# Magnetic Field Structure in Spheroidal Star-Forming Clouds

**Philip C. Myers[1],  Shantanu Basu[2], and Sayantan Auddy[1,2]**

[1]Harvard-Smithsonian Center for Astrophysics, 60 Garden Street,  Cambridge MA 02138 USA

[2]Department of Physics & Astronomy, University of Western Ontario, London, ON, Canada, N6A 3K7

**Abstract.**  A model of magnetic field structure is presented to help test the prevalence of flux freezing in star-forming clouds of various shapes, orientations, and degrees of central concentration, and to estimate their magnetic field strength. The model is based on weak-field flux freezing in centrally condensed Plummer spheres and spheroids of oblate and prolate shape. For a spheroid of given density contrast, aspect ratio, and inclination, the model estimates the local field strength and direction, and the global field pattern of hourglass shape. Comparisons with a polarization simulation indicate typical angle agreement within 1 - 10 degrees. Scalable analytic expressions are given to match observed polarization patterns, and to provide inputs to radiative transfer codes for more accurate predictions. The model may apply to polarization observations of dense cores, elongated filamentary clouds, and magnetized circumstellar disks.

*keywords:* ISM: clouds—stars: formation

# 1. Introduction

## 1.1. Magnetic fields and star formation

Magnetic fields are considered an important part of the star formation process (Mestel & Spitzer 1956; Mouschovias 1976; Galli & Shu 1993; Padoan & Nordlund 1999; McKee & Ostriker 2007; Kudoh & Basu 2008; Crutcher 2012; H. Li et al. 2014; Z.-Y. Li et al. 2014). Some key questions include (1) how important is the field energy density compared to that of turbulence, self-gravity, and thermal motions, in early and late stages of star formation? (2) how does flux freezing relax to allow disk and star formation? and (3) how do the spatial structure of the gas density and the magnetic field relate?

Addressing these questions requires improvement in our knowledge of the magnetic field in star-forming gas. Progress in measuring the magnetic field strength has been slow, due to limitations in available techniques (Crutcher 2012). Determination of the line-of-sight component of the field strength, measured by the Zeeman effect, has been limited by observational sensitivity and resolution (e.g. Crutcher et al. 2010). Estimation of the mean field strength in the plane of the sky from the dispersion in directions of dust polarization (Davis 1951; Chandrasekhar & Fermi 1953; DCF) requires enough measurements to separate the turbulent and ordered contributions (Hildebrand et al. 2009, Kandori et al. 2017). The comparison of spectral line widths of similar ionic and neutral molecular species (Houde et al 2000, Li & Houde 2008) is limited by the requirement that the neutral and ionic tracers be cospatial. Field strength estimates by Faraday rotation (Wolleben & Reich 2004) are limited to ionized gas which does not coincide with star-forming gas. Estimates based on the gradient in the centroid of turbulent spectral lines (Gonzalez-Casanova & Lazarian 2017, Yuen & Lazarian 2017) apply primarily to the low-density gas surrounding star-forming clouds.

## 1.2. Magnetic polarization

Observations of ordered polarization due to magnetically aligned grains may provide a way forward, since its structure contains information independent of the statistical fluctuations analyzed by the DCF technique. Pioneering polarization observations of OMC-1 at 100 $\mu$m and 350 $\mu$m showed evidence of the hourglass polarization pattern expected in simple models of flux freezing (Schleuning 1998). Submillimeter observations with the JCMT SCUPOL instrument (Matthews et al. 2009) and with the *Planck* satellite (*Planck* Collaboration I. 2016) show highly ordered polarization directions in and around numerous star-forming clouds. The *Planck* polarization directions agree well with those of near-infrared observations of the same regions. This agreement supports the idea of magnetic grain alignment in star-forming regions, on scales of a few 0.1 pc to a few 10 pc (Soler et al. 2016).

At the same time the large-scale structure of nearby star-forming clouds has become available in much greater detail, due to imaging in near-infrared dust extinction by wide-field array cameras (e.g. Lombardi et al. 2006) and due to imaging in far-infrared dust emission by the *Herschel* satellite (e.g. André et al. 2010).

Ordered polarization is seen in the Musca, B211, and L1506 filamentary dark clouds at 1 mm wavelength, where *Planck* polarization directions lie within 10 deg of perpendicular to the crest direction in the denser Musca and B211 filaments, while they are parallel to the crest direction in the less dense filament L1506 (Planck Collaboration Int. XXXIII 2016). Similarly, polarization directions are observed to be mostly perpendicular to the massive infrared dark cloud filament G11.11-0.12 (Pillai et al. 2016).

Recent improvements in the sensitivity and resolution of polarization measurements at far-infrared and submillimeter wavelengths offer new opportunities to relate detailed maps of dust polarization to corresponding maps of dust column density. These include the balloon-borne BLAST-Pol mission (Fissel et al. 2016) and the JCMT BISTRO survey (Pattle et al. 2017), which have made polarization maps of large-scale clouds and filaments. Along with the SAO Submillimeter Array (SMA; Blundell 2007), ALMA (Cox 2016), and SOFIA (Zinnecker et al. 2015) which provide finer angular resolution, these facilities offer an order of magnitude increase in the typical number of independent polarization directions in a map, compared to earlier observations. This increase enables more detailed matching between models and observations, and better discrimination among models.

An increasing number of sources have been reported as examples of polarization arising from flux freezing in star-forming clouds. Dense cores showing high-resolution hourglass submillimeter polarization patterns include NGC1333 IRS4 (Girart et al. 2006, Frau et al. 2011); IRAS 16293 A (Rao et al. 2009); L1157-mm (Stephens et al. 2013); and B335 (Maury et al. 2018). More massive cores with hourglass patterns include G31.41+0.31 (Girart et al. 2009), W51e2 (Tang et al. 2009), and G240.31+0.07 (Qiu et al. 2014). In addition, FeSt 1-457 (Kandori et al. 2017) and possibly L1544 (Clemens et al. 2016) show hourglass patterns at shorter wavelengths. Among these hourglass sources, NGC1333 IRS4 and B335 have been compared quantitatively to predictions of simulations based on models assuming non-ideal magnetohydrodynamic (henceforth MHD) conditions (Goncalves et al. 2008, Frau et al. 2011, Masson et al. 2016, Maury et al. 2018).

Circumstellar disks are expected to be magnetized (Z.-Y. Li et al. 2014), and if so they may have detectable polarization due to magnetically aligned grains in a disk geometry. Such magnetized grain polarization has been suggested to explain the patterns observed with ALMA in protostellar envelopes or outer disks in Perseus (Cox et al. 2018), Kwon et al. 2018), Ophiuchus (Sadavoy et al. 2018), and in the Pipe Nebula (Alves et al. 2018). If such polarization can be distinguished from other simultaneously acting mechanisms of polarization, including dust scattering, radiative alignment, and mechanical alignment described in Andersson et al. (2015), it may provide useful constraints on magnetized disk physics. Observed polarization patterns may also be too complex to reveal a simple ordered structure. For example, ALMA polarization observations of the environment of the protostar Ser-emb-8 suggest that turbulent motions may dominate over its magnetic structure (Hull et al. 2017).

Thus it seems useful to describe the flux-frozen magnetic field structure expected for simple density models, to allow comparison with well-sampled polarization patterns in star-forming regions. Such a comparison can test the applicability of flux freezing in regions of low and high mass star formation, in isolated and clustered settings, and in regions where turbulent and ordered magnetic fields have comparable influence. It can further test the extent and role of flux freezing in regions whose geometry is not spherical, including prolate dark cloud filaments and oblate circumstellar disks.

## 1.3. Magnetic field structure in spheres and spheroids

Flux freezing has previously been discussed in oblate spheroids by Strittmatter (1966; S66) and in flattened layers by Nakano & Nakamura (1978; NN78). Each of these papers found a critical mass-to-flux ratio for gravitational instability. A virial analysis indicates that the critical mass-to-flux ratio for the spheroid differs from that of a sphere by a factor of at most 1.9, when the spheroid is a highly flattened disk with eccentricity approaching 1 (S66). Similarly, the stability of an isothermal layer threaded by a uniform field was studied with a perturbation analysis by NN78. They concluded that perturbations become unstable when the mass-to-flux ratio lies within the range of critical values found by S66.

S66 and NN78 did not obtain magnetic field structure due to flux freezing in a centrally condensed body, as did Mestel (1966; M66) for the sphere, and as this paper does for spheroids. Numerical simulations have obtained the expected polarization structure for the collapse of an initially magnetized sphere, which becomes more oblate as the collapse proceeds (e.g., Kataoka et al. 2012; K12; Vaytet et al. 2018). However these simulations do not provide an analytic form suitable for fitting to observations.

M66 obtained magnetic field structure by assuming mass and flux conservation of a sphere of uniform density and field strength, as it is compressed with spherical symmetry into a centrally concentrated sphere. The associated magnetic field has a "primary distortion" structure computed solely from flux freezing and mass conservation in the specified density profile. The compressed sphere is out of equilibrium, since it has unbalanced magnetic forces directed toward the equator. M66 also compute a "secondary distortion" structure, where the gas responds to both magnetic and gravitational forces, increasing its density in the equatorial plane. The resulting disequilibrium structure is sometimes called a "pseudo-disk" (Galli & Shu 1993).

The "primary distortion" is thus a first approximation to the magnetic field structure in a condensation. The approximation is most accurate when magnetic forces are sufficiently weak compared to gravity, as shown in the simulation comparison in Section 3.5.

## 1.4. This model

This "Spheroid Flux Freezing" (SFF) model computes the primary-distortion magnetic field structure in the plane of the sky for spheres and for spheroids oriented parallel and perpendicular to the direction of the magnetic axis.

In this model, Plummer spheroids represent axisymmetric, centrally condensed structures with spherical, prolate, or oblate geometry, as described in Myers (2017). Their magnetic flux structure is predicted, based on flux freezing and mass conservation as in M66 and Mestel and Strittmatter (1967). Concentric flux tubes are sampled in meridian planes of the spheroid. Expressions are given to account for inclination of the magnetic axis from the plane of the sky, and for the effect of integration along the line of sight. When the model fits the density and polarization structure of a condensation, it provides estimates of the magnetic field structure.

The analytic expressions for SFF field line shapes and directions apply to a greater variety of cloud morphology than the spheres studied by M66, and their simple mathematical form is easier to use than some earlier analytic descriptions (Galli & Shu (1993a,b); Ewertowski & Basu (2013). This work is more approximate than numerical simulations of magnetized star-forming regions, (e.g. Machida et al. 2008, Kudoh & Basu 2011), but its analytic formulation provides physical insight and allows fitting to observed polarization patterns. The flux structures derived from the present model can be used as input to polarization radiative transfer codes such as that of Reissl et al. (2014) or Padovani et al. (2012).

## 1.5. This paper

In this paper, Section 2 describes the assumptions, definitions, and basic equations used to obtain the structure of a flux tube of a given flux, and the value of the polarization angle in the plane of the sky at any map point. Section 3 gives spatial patterns on the plane of the sky, showing their variation with the ratio of peak to background density, with the spheroid aspect ratio, and with the inclination of the magnetic axis from the projected symmetry axis. This section also compares SFF field directions to polarization directions, in simulations and observations. Section 4 describes inference of magnetic field strength from the SFF model.

Section 5 discusses the results, their limitations, and possible applications. Section 6 summarizes the paper and gives conclusions.

## 2. Magnetic field structure

### 2.1. Flux freezing model

2.1.1. Flux in terms of radius and mean density. The SFF model assumes that a spheroidal volume in a medium of uniform density, uniform field strength, and uniform field direction condenses into a centrally concentrated spheroid of the same mass and shape, threaded by the same flux. Similar idealized initial conditions were assumed in studies by Spitzer (1968), Mouschovias (1976 *a,b*), and Mouschovias & Spitzer (1976). In particular, Mouschovias (1976 *a,b*) calculated the exact magnetohydrostatic equilibria that could be obtained by continuous deformation of gas and field lines when starting from a uniform spherical state. The final equilibria resembled oblate spheroids.

The present treatment extends that of M66 from spheres, assumed to have a Gaussian density profile, to spheroids, assumed to have a Plummer density profile. The spheroid shape can be spherical, oblate, or prolate. Unlike the sphere, oblate and prolate spheroids each have a single axis of rotational symmetry. This axis is assumed to lie either parallel or perpendicular to the magnetic axis, whence the spheroid is here called "parallel" or "perpendicular." The field line shapes are set by the shapes of their enclosing flux tubes, since each flux tube is a stream function for its field lines (Spruit 2016).

For spheroids of spherical shape and for parallel spheroids, the horizontal cross-section shape of each flux tube is a circle centered on the magnetic axis, defined as the $z$-axis. At each height $z$, the radius of the circle is set by the flux, and by the intersection of the $x$ - $y$ plane with a surface of constant density of the spheroid.

For perpendicular spheroids, the flux tube cross section shape is an ellipse having the same orientation as the spheroid. For a perpendicular prolate spheroid each flux tube ellipse has major axis in the long-axis direction; for a perpendicular oblate spheroid each flux tube ellipse has minor axis in the short-axis direction. These relations follow from equations (5), (11), and (12), and are shown schematically in Figure A-1.

The vertical structure of the flux tube is obtained from mass and flux conservation between initial and final structures of the same geometrical shape. When the magnetic and spheroid axes are parallel, let $\Phi_c$ be the flux through a circle of cylindrical radius $r_c$, defined by the polar angle $\theta_c$ from the $z$ - axis. The corresponding initial flux $\Phi_{ci}$ is enclosed by a circle of greater cylindrical radius $r_{ci}$ in the initial spheroid at the same polar angle. Similarly, let $\Phi_e$ be the flux enclosed by an ellipse of semi-major axis $r_e$, making polar angle $\theta_e$ with the $z -$ axis, with corresponding initial flux $\Phi_{ei}$ enclosed by an ellipse of semi-major axis $r_{ei}$ at the same polar angle. Henceforth the subscripts $c$ and $e$ indicate regions whose flux tubes have circular or ellipsoidal cross section.

The initial flux through these flux tubes can be written

$$\Phi_{ci} = \pi r_{ci}^2 B_u \qquad\qquad (1)$$

and

$$\Phi_{ei} = \left(\frac{\pi}{D}\right) r_{ei}^2 B_u \qquad\qquad (2)$$

where $B_u$ is the initial uniform field strength. In equation (2), $D$ is the axis ratio of the oblate or prolate spheroid whose symmetry axis is perpendicular to the magnetic axis.

In this model, the initial uniform density $n_u$ is equal to the final density at large distance from the spheroid center, which is here called the "background" density. Similarly, the initial uniform field strength $B_u$ is equal to the final "background" field strength. These uniform background quantities are analogous to the constant-pressure gas which truncates an isothermal Bonnor-Ebert (BE) sphere (Bonnor 1956; Ebert 1955), as discussed in Section 4

The final fluxes $\Phi_c$ and $\Phi_e$ are obtained in terms of the background field strength and the mean density of the final spheroid. Mass conservation between the initial and final spheroid gives $r_{ci} = r_c \bar{v}^{1/3}$ in equation (1) and $r_{ei} = r_e \bar{v}^{1/3}$ in equation (2). Here $\bar{v} \equiv \bar{n}/n_u$ is the mean density whose bounding surface defines $r_c$ or $r_e$, normalized by the background density. Conservation of the flux between an initial and final flux tube gives $\Phi_c = \Phi_{ci}$ and $\Phi_e = \Phi_{ei}$, whence equation (1) becomes

$$\Phi_c = \pi r_c^2 \bar{v}^{2/3} B_u \qquad (3)$$

and equation (2) becomes

$$\Phi_e = \left(\frac{\pi}{D}\right) r_e^2 \bar{v}^{2/3} B_u \quad . \qquad (4)$$

Equations (3) and (4) indicate that the flux varies as the 2/3 power of the mean density $\bar{n}$ for both spheres and spheroids which contract with constant shape. This property follows only for clouds whose fields are too weak to significantly impede motions across field lines (Crutcher 2012). The three associated SFF features (1) field energy weaker than gravity and stronger than turbulence, (2) mean field strength scaling as $\bar{n}^{2/3}$ for dense gas, and (3) core maps both round and elongated, are also seen in simulations 2 and 3 of Mocz et al. (2017).

2.1.2. Alternate model. A different way to describe the structure of a flux tube of cylindrical radius $r_c$ obtains the vertical component of the field strength $B_z$ at cylindrical radii $r_c' \leq r_c$ and height $z$ from the above flux freezing model. Then one can integrate $B_z$ over the enclosed area, i.e. $\Phi(r_c, z) = 2\pi \int_0^{r_c} dr_c' \, r_c' B_z(r_c', z)$. Expressions for $B_z(r_c, z)$ in terms of the condensation density can be obtained by differentiating the flux as in M66, or by approximating $B_z$ as $B_z \approx B_u \bar{v}^{2/3}$. However, the flux expressions in equations (3) and (4) are simpler to evaluate analytically, so they are used in the following calculations.

2.1.3. Dynamical assumptions. The SFF model describes a centrally concentrated axisymmetric spheroid which has condensed from an initially uniform magnetized medium while conserving flux, mass, and shape. The spheroid is not required to be in MHD force balance, but it is expected to be close to self-gravitating equilibrium in at least one dimension, since the Plummer sphere density as a function of spherical radius is similar to that of the isothermal sphere (Bonnor 1966; Ebert 1965). Also, an extended Plummer prolate spheroid has density as a function of cylindrical radius similar to that of a self-gravitating polytropic cylinder (Toci & Galli 2015).

The ordered magnetic field energy is assumed to be consistent with observations of star-forming regions discussed in Section 1. It is therefore assumed to be stronger than the turbulent magnetic or turbulent kinetic energy, so that the field structure is dominated by its ordered component. The field energy is also assumed to be weaker than the gravitational energy of the condensation, so that the condensation can fragment, collapse, and form stars. The configuration has unbalanced magnetic forces toward its equator, as described in Section 4. It is further assumed that these forces are weak enough so that the spheroidal description remains useful.

## 2.2. Spheroid density model

This section presents a density model which describes spheres, prolate spheroids, and oblate spheroids embedded in a uniform background medium. Each spheroid is presented in two perpendicular orientations.

2.2.1. Spheroid density. The adopted condensation model can approximate spherical cores, elongated filaments, and flattened shells and disks. Its density decreases monotonically toward a uniform background from a finite local maximum. It is based on a Plummer spheroid embedded in a uniform medium,

$$\nu = 1 + \nu_0[1 + \omega^2]^{-p/2} \ , \tag{5}$$

where $\nu \equiv n/n_u$ is the density normalized to the background value, $\nu_0 \equiv n_0/n_u \geq 1$ is the peak spheroid density normalized to the background value, and $\omega \equiv [(\xi/A)^2 + (\eta/B)^2 + (\zeta/C)^2]^{1/2}$ is the "spheroid radius," or the dimensionless distance from the center to a point where the density is $n$. The dimensionless coordinates $\xi, \eta,$ and $\zeta$ are respectively $x$, $y$, and $z$, each normalized to the scale length $r_0 \equiv \sigma/\sqrt{4\pi Gmn_0}$ . Here $\sigma$ is the one-dimensional thermal velocity dispersion, $G$ is the gravitational constant, and $m$ is the mean particle mass 2.33 $m_H$. The index $p$ is set to 2 as in some dense core models which approximate a Bonnor-Ebert sphere (Tafalla et al. 2004) or as in models of centrally condensed filaments (Arzoumanian et al. 2011). Variation of $p$ causes slight changes in the resulting flux tube shape, but does not alter the basic results.

The axis ratios $A$, $B$, and $C$ specify the spheroid shape. The shapes having flux tubes of circular cross section are the sphere, where $A = B = C = 1$, and the "parallel spheroids." Each of these parallel spheroids has its symmetry axis parallel to the magnetic axis, defined here to be the $z$ - axis. They are the parallel prolate spheroid, with $C > A = B = 1$, and the parallel oblate spheroid, with $A = B > C = 1$.

The spheroids having flux tubes of ellipsoidal cross section are "perpendicular spheroids," i.e. each of these has its symmetry axis perpendicular to the magnetic axis. These symmetry axes are chosen to lie in the $x$ - direction, for a perpendicular prolate spheroid with $A > B = C = 1$; and for a perpendicular oblate spheroid with $B = C > A = 1$.

2.2.2. Spheroid mean density. The mean density in equations (3) and (4) is obtained by integration of equation (5) with $p = 2$, according to $\bar{\nu} = (3/\omega^3)\int_0^\omega d\omega' \omega'^2 \nu(\omega')$. Then the spheroid whose normalized density is $\nu$ has normalized mean density $\bar{\nu}$, given by

$$\bar{\nu} = 1 + \frac{3\nu_0}{\omega^2}\left(1 - \frac{\tan^{-1}\omega}{\omega}\right) \qquad . \tag{6}$$

The spheroid radius $\omega$ depends on density $\nu$ and peak spheroid density $\nu_0$ as

$$\omega^2 = \frac{\nu_0}{\nu - 1} - 1 \qquad\qquad (7)$$

from equation (5). Equation (7) shows that $\omega$ is constant over a surface of constant density $\nu$. For a given spheroid shape, surfaces of increasing $\omega$ are concentric spheroids.

Expressions for $\omega^2$ in terms of coordinates follow from equation (5) according to the shape and orientation of the spheroid. For the sphere, $\omega^2$ is given by

$$\omega^2 = \xi^2 + \eta^2 + \zeta^2 \quad . \qquad\qquad (8)$$

In equation (8) and in the following equations, the scale factor $A$, $B$, or $C$ is shown explicitly only if it exceeds unity. For the parallel prolate spheroid,

$$\omega^2 = \xi^2 + \eta^2 + (\zeta/C)^2 \quad . \qquad\qquad (9)$$

For the parallel oblate spheroid,

$$\omega^2 = (\xi/A)^2 + (\eta/A)^2 + \zeta^2 \quad . \qquad (10)$$

For the perpendicular prolate spheroid,

$$\omega^2 = (\xi/A)^2 + \eta^2 + \zeta^2 \qquad , \qquad (11)$$

and for the perpendicular oblate spheroid,

$$\omega^2 = \xi^2 + (\eta/C)^2 + (\zeta/C)^2 \quad . \qquad (12)$$

### 2.3. Flux tube structure

To compute contours of constant flux in the $x$ - $z$ plane, $\eta = 0$ is assumed in equations (8) - (12). Then dimensionless versions of equations (3) and (4) are used to relate the flux to the coordinates of the spheroid. Henceforth dimensionless quantities are referred to without the "dimensionless" prefix. For flux tubes of circular cross section, the cylindrical radius is $\xi_c \equiv x_c/r_0$ for the sphere, the parallel prolate spheroid, and the parallel oblate spheroid. For flux tubes of ellipsoidal cross section, the cylindrical radius in the $x$ - direction is $\xi_e \equiv x_e/r_0$, which equals the semi-major axis for the perpendicular prolate spheroid, or the semi-minor axis for the perpendicular oblate spheroid.

The flux $f$ through each spheroid is normalized by $\Phi_0 = \pi r_0^2 B_u$, the initial field strength times the area of a circle whose radius equals one spheroid scale length. Then each flux $f_c = \Phi_c/\Phi_0$ and $f_e = \Phi_e/\Phi_0$ is simply related to the coordinates of a point on a flux tube enclosing that flux, using equations (3), (4), and (6).

For spheres, parallel prolate spheroids, and parallel oblate spheroids, a flux tube of flux $f_c$ has cylindrical radius $\xi_c$ in the $x$ - direction and spheroid radius $\omega$ in the $x$ - $z$ plane related by

$$\xi_c = f_c^{1/2} \left[ 1 + \frac{3\nu_0}{\omega^2} \left( 1 - \frac{\tan^{-1}\omega}{\omega} \right) \right]^{-1/3} \qquad (13)$$

based on equations (3) and (6).  For a spherical condensation the height $\zeta_c$ of a flux tube point with this radius $\xi_c$ is

$$\zeta_c = (\omega^2 - \xi_c^2)^{1/2} \qquad (14)$$

based on equation (8).  The height for the parallel prolate spheroid is

$$\zeta_c = C(\omega^2 - \xi_c^2)^{1/2} \qquad (15)$$

based on equation (9).  The height for the parallel oblate spheroid is

$$\zeta_c = [\omega^2 - (\xi_c/A)^2]^{1/2} \qquad (16)$$

based on equation (10).

For perpendicular prolate and oblate spheroids, a flux tube of flux $f_e$ has cylindrical and spheroid radii related in the same way as in equation (13), via equations (4) and (6),

$$\xi_e = f_e^{1/2} \left[ 1 + \frac{3\nu_0}{\omega^2} \left( 1 - \frac{\tan^{-1}\omega}{\omega} \right) \right]^{-1/3} . \qquad (17)$$

For the perpendicular prolate spheroid, the corresponding height is

$$\zeta_e = [\omega^2 - (\xi_e/A)^2]^{1/2} \qquad (18)$$

based on equation (11). For the perpendicular oblate spheroid, the corresponding height is

$$\zeta_e = C(\omega^2 - \xi_e^2)^{1/2} \qquad (19)$$

based on equation (12).

Equations (13)-(19) give simple relations for the shape of a flux tube in the $x - z$ plane for a flux-frozen, $p = 2$ Plummer spheroid oriented parallel or perpendicular to the initial magnetic field. These relations depend only on the assumed flux value $f_c$ or $f_e$, the peak density ratio $\nu_0$, the spheroid aspect ratio $A$ or $C$, and its orientation. In these equations the spheroid radius $\omega$ serves as a dummy variable, increasing from 0 to $\infty$ as $\nu$ decreases from $\nu_0 + 1$ to 1.

## 2.4. Projected flux tube structure

A full description of the observable polarization due to magnetic grain alignment requires integration of the polarized and unpolarized emission along the line of sight, taking into account the polarized scattering and emission properties of the dust grains and their radiative transfer, as in the simulations of Goncalves et al. (2008), Padovani et al. (2013), K12, or Reissl et al. (2014). Such a detailed numerical treatment is beyond the scope of this paper. Instead the present analytic approach is scalable, to allow more accurate fitting of observed polarization directions, and to allow estimation of field strength and density contrast.

2.4.1. Flux profiles. Here the line-of-sight average shape of each 3D flux tube is approximated by a 2D "flux profile." This profile represents a flux tube by projection onto the plane of the sky of at least one planar cut through the tube. The intersection of concentric flux tubes with the plane of the sky is the standard way to visualize field lines in 2D when the flux tube axis also lies in the plane of the sky (Mestel & Strittmatter 1967, Mouschovias 1976a, Galli & Shu 1993, Spruit 2016).

This single-plane representation becomes incomplete when a flux tube is inclined through a polar angle $\theta > 0$ about a horizontal axis in the plane of the sky. In that case, curved field lines in the front and rear of the tube have "front-back asymmetry," or different projected directions at intermediate inclinations (K12, figure 7). These different directions combine in a weighted average when the radiative transfer of the polarized flux is computed numerically. This effect cannot be represented analytically by sampling a flux tube in a single plane.

A more accurate analytic approximation takes into account field lines in the front and rear of the flux tube. The field structure is first computed in a flux tube having inclination $\theta = 0$ about the vertical axis, in planes passing through the axis at different azimuth angles $\phi$. Then the field structure $\zeta(\xi)$ in each plane is used with $\phi$ to calculate the projected structure when the tube is inclined through $\theta > 0$.

In this case, let a circular flux tube with axis along $\theta = 0$ have cylindrical radius $\xi_c$ at vertical height $\zeta_c$ as in Section 2.3 above, and let a point with these coordinates be further specified by the azimuth angle $\phi$. After inclination of the tube through $\theta > 0$ the horizontal and vertical coordinates $\xi_{\theta\phi}$ and $\zeta_{\theta\phi}$ are related to the original coordinates by

$$\xi_{\theta\phi} = \xi_c \cos\phi \qquad\qquad (20)$$

and

$$\zeta_{\theta\phi} = \zeta_c \cos\theta - \xi_c \sin\phi \sin\theta \quad . \qquad (21)$$

Equations (20) and (21) can be obtained by simple trigonometry when the angles are viewed in the $y$ - $z$ plane, as illustrated in Figure A2 of the Appendix. They can also be generalized to obtain the structure of an inclined flux tube of ellipsoidal cross section.

2.4.2. Selection of azimuth planes. Equations (20) and (21) can be used with multiple azimuth samples of a flux tube to describe inclined flux tube structure, for detailed comparison with observations (e.g. Sadavoy et al. 2018). Here just two perpendicular planes are discussed. The planes $\phi = 0$ to $\pi$ and $\phi = \pi/2$ to $3\pi/2$ are the simplest choice, because then the front-back asymmetry occurs only in the plane $\phi = \pi/2$ to $3\pi/2$, which is viewed edge-on.

Application of equations (20) and (21) for the plane from $\phi = 0$ to $\pi$ gives a projected pattern with a tilted hourglass shape. Applying them for the perpendicular plane from $\phi = \pi/2$ to $3\pi/2$ gives a vertical component which lies on the projected symmetry axis of the hourglass. The hourglass component, denoted with subscript $h$, has four quadrant curves with reflection symmetry about the $\xi$ - and $\zeta$ - axes,

$$\xi_h = \pm \xi_c \qquad\qquad (22)$$

$$\zeta_h = \pm \zeta_c \cos \theta \qquad . \qquad (23)$$

The vertical component, denoted with subscript v, is a superposition of four vertical lines on the $\zeta$ – axis,

$$\xi_v = 0 \qquad\qquad (24)$$

$$\zeta_v = \pm \zeta_c \cos \theta \pm \xi_c \sin \theta \qquad . \qquad (25)$$

Each straight line in equations (24) - (25) is an edge-on view of a curved line in equations (22) - (23), whose height has been increased or decreased by the vertical increment $\xi_c \sin \theta$ due to front-back asymmetry. The average over this asymmetry is $\overline{\zeta_\nu} = \pm \zeta_c \cos\theta$ for most of the range of $\theta$. This average changes to $\overline{\zeta_\nu} = \pm \xi_c \sin\theta$ when $\theta$ approaches $\pi/2$, depending on which term dominates in equation (25). Examples of inclined flux tubes and further details are given in Sections 3.2 and 3.4.

2.4.3. Selection of flux values. The model magnetic field lines should approach uniform spacing and direction at large distance $\omega \gg 1$ from the condensation center, where the field strength approaches the initial uniform value $B_u$. Also, at $\omega \gg 1$ the flux tube radii $\xi$ approach $f^{1/2}$ according to equations (13) and (17). This property implies that a sequence of concentric flux tubes should have flux values which increase as a sequence of squared integers, i.e. the flux of the $i$th flux tube should follow $f_i = i^2 f_1$ where $i = 1,2,3,...$

With this prescription, field lines at large radius have uniform linear spacing $\Delta x = r_0 f_1^{1/2}$ set by the choice of the flux $f_1$ of the innermost flux tube in the model. To match the spatial sampling $\Delta x_{obs}$ of a polarization observation, one can choose $f_1 = (\Delta x_{obs}/r_0)^2$. In the examples of Section 3, $f_1$ is chosen to show a substantial variation of hourglass shape with increasing peak density. This is done by requiring the waist of the innermost flux tube to have a radius of one scale length in the $x$-direction, or $Ar_0$. Here $A = 1$ for the sphere, the parallel prolate spheroid, and the perpendicular oblate spheroid, and $A > 1$ for the parallel oblate spheroid and the perpendicular prolate spheroid. Then $f_1 = A^2[1 + 3\nu_0(1 - \tan^{-1} 1)]^{2/3}$ using equation (13) with $\eta = \zeta = 0$. For fiducial peak densities $\nu_0 = 30, 300,$ and $3000$, the resulting field line spacing is $\Delta x/Ar_0 = 2.73, 5.79,$ and $12.5$.

2.4.4. Magnetic field direction at any map position. For a given spheroid model, the foregoing procedures yield continuous contours of magnetic field direction, projected on the plane of the sky, from the structure of their associated flux tubes. These contours are useful to image the global structure of the magnetic field, and to determine whether the observed polarization pattern is broadly consistent with flux freezing.

However, the curved field lines for a set of flux values cannot pass exactly through all the points in a polarization map, which are typically arranged in a uniform grid. This departure limits the point-by-point comparison of observed and model polarization angles, which is useful to evaluate the goodness of fit of a global polarization model, and to optimize model parameters (e.g. Frau et al. 2011, Maury et al. 2018, Alves et al. 2018).

For such point-by-point comparison it is preferable to model the local polarization angle at each map point, but it is not necessary to compute the global structure of the flux tube associated with that point. While the "global" calculation uses the equation between the constant flux and the flux tube structure as in equation (13), the "local" calculation uses the derivative of the same equation. Then the derivative $df(\xi, \zeta)/d\xi$ expresses the field line slope $d\zeta/d\xi$ as a function of coordinates and spheroid density contrast, but independent of the flux.

At a given map point the polarization position angle $\theta_{pol}$ can be written $\theta_{pol} = \theta_0 + \theta_B$ where $\theta_0$ is the map reference direction in the plane of the sky (*e.g.*, celestial north). Here the local magnetic field polar angle $\theta_B$ is defined by the ratio of field components, $\theta_B = \tan^{-1} B_x/B_z$, or equivalently by the slope of a contour of constant flux, $\theta_B = \tan^{-1} d\xi_c/d\zeta_c$. The same expression for $\theta_B$ has been obtained from the field components in M66 equations (11) and (12) after conversion from polar to cartesian coordinates, and from the constant flux contour equations in Section 2, using the identity $d\bar{v}/d\omega = (3/\omega)(v - \bar{v})$, giving

$$\theta_B = \tan^{-1}\left[\frac{1-t}{s(s^{-2}+t)}\right] \qquad (26)$$

where $s \equiv \xi/\zeta$ and $t \equiv v/\bar{v}$.

Equation (26) gives field directions expected for hourglass structure in a uniform background. These directions approach the model vertical direction at the equator and at large distance from the center, and they approach a maximum value at intermediate directions. The predicted field line structure reflects the differing dependence on density of the polar and radial magnetic field components derived in M66. Equation (26) is applicable to any spheroidal density model, including the *p = 2* Plummer spheroid used in this paper as an example.

The derivation of equation (26) in two ways,  from the ratio of magnetic field components at a point in a meridian plane, and from the slope of the curve defined by the intersection of the plane and the flux tube which passes through that point, demonstrates that the SFF model has the expected equivalence of magnetic field line direction and flux tube slope (M66, Mouschovias 1976a, Spruit 2016).

On the other hand, if the basic assumption of  contraction at constant shape were relaxed, the field structure would depart from that in equation (26) and its morphology would change with time.  For a spherically symmetric initial volume contracting preferentially along vertical field lines, the condensation would become progressively more oblate.  If this structure were approximated as an oblate spheroid whose aspect ratio and peak density each increase with time, then for weak fields equation (26) could still be used, with the appropriate time-dependent values of $A$ and $v_0$.  In this case the field structure would resemble an hourglass shape whose pinched zone extends progressively farther outward in the equatorial plane, from an initial shape resembling that in  Figure 1b toward a final shape resembling that in Figure 5a.

In matching observed polarization patterns, it may be desirable to first optimize model parameters with point-by-point use of equation (26), and then to apply these parameters to compute a global field pattern following the procedure given earlier in this section.

## 3.  Magnetic Field Patterns

This section shows properties of flux-frozen magnetic field patterns calculated according to Section 2 above.  These properties include variation of the hourglass shape for Plummer spheres of increasing peak density in Section 3.1, for spheres with increasing inclination in Section 3.2, for perpendicular and parallel prolate spheroids of increasing peak density in Section 3.3, and for parallel oblate spheroids of increasing inclination, in Section 3.4. Section 3.5 presents a comparison of SFF field directions with polarization angle directions in a numerical simulation of a collapsing BE sphere (K12).  Typical angle deviations between SFF field directions and polarization directions are shown to be $\sim$ 10 deg or less, due mainly to line-of-sight integration of field directions.  A simple modification to the SFF evaluation plane reduces this deviation to less than ~1 deg. Section 3.6 presents a comparison of SFF field directions with polarization angles observed in the environment of the protostar VLA 1623A, based on ALMA observations (Sadavoy et al. 2018).

### 3.1. Plummer spheres with increasing peak density

Figure 1 shows flux tubes in the plane of the sky for $p = 2$ Plummer spheroids with normalized density ratio $\nu_0 = 30$, 300, and 3000, calculated for the two planes $\phi = 0$ to $\pi$ and $\phi = \pi/2$ to $3\pi/2$ as described in Section 2.4.2 above. Each flux tube coordinate is normalized by the spheroid scale length $r_0$. In each figure the scale bar is computed according to the assumed values $T = 10$ K, as is typical for dense gas in nearby star-forming regions (Rosolowsky et al. 2008), and according to a fiducial background density $n_u = 300$ cm$^{-3}$ and background field strength $B_u = 10\,\mu$G as in Crutcher (2012).

Each plot in Figure 1 shows a central vertical line, which is the projection of the field lines in the plane perpendicular to the plane of the sky. Each plot also shows seven hourglass field lines in each quadrant, calculated from the flux tube coordinate equations (13) and (14). The flux $f_1$ of the innermost flux tube is chosen so that its waist passes within one scale length of the spheroid center. The flux of the $i$th flux tube is set by $f_i = i^2 f_1$, $i = 1, ...7$ as described in Section 2.4.3.

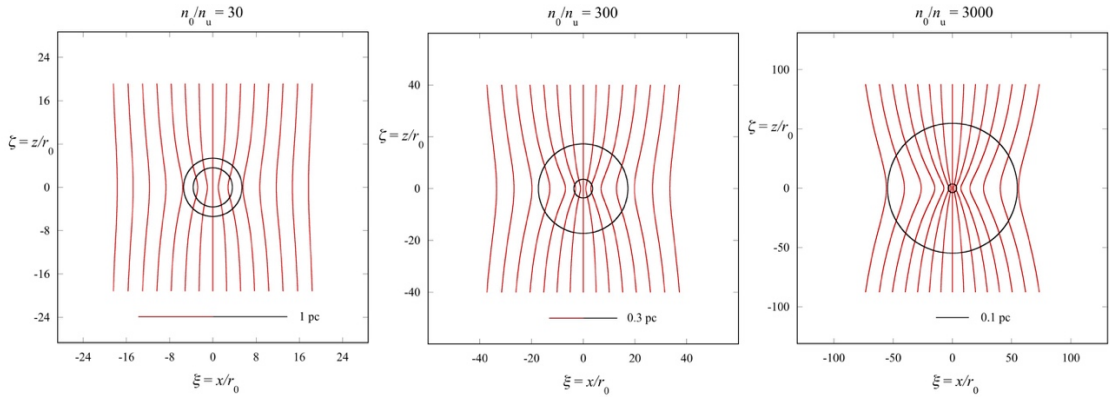

**Figure 1.** Flux tube lines associated with a $p = 2$ Plummer sphere whose peak density exceeds the background density by factors of 30 (*left*), 300 (*center*), and 3000 (*right*). The $i$th line from the center line marks the intersection of the $x$ - $z$ plane with a flux tube of dimensionless flux $i^2 f_1$, where $f_1$ is defined in Section 2.4.3. The 0.3 pc scale bar is based on background density $n_u = 300$ cm$^{-3}$. The inner circle marks a density less than the peak value by a factor 14.0, as in a critical Bonnor-Ebert sphere (Galli et al. 2002, McKee & Ostriker 2007, hereafter MO07). The outer circle marks a density with twice the background value.

For comparisons, each panel in Figure 1 shows a measure of the condensation size as a contour of constant density at a value $n_0/14.0$, the density at the boundary of a critically stable pressure-bounded sphere (Bonnor 1956; Ebert 1955; Galli et al. 2002). The outer circle marks the transition to the background density with a contour of constant density $2n_u$. Each of these circles represents the shape of an isodensity contour without regard to magnetic forces, on the assumption that the field is sufficiently weak.

In a more realistic estimate, the isodensity surface becomes more oblate as gas flows preferentially along field lines. However for weak fields, early times, and large scales this effect is modest. In a simulated collapse of a BE sphere initially threaded by a uniform field, with mass-to-flux ratio 6.6, the aspect ratio of the isodensity contour in the plane of the sky at half the initial radius and at twice the free-fall time is 1.05, only slightly different from the aspect ratio of unity for the initial circular shape (K12, Figure 4b). This representation of isodensity contours in the weak-field limit is followed in all subsequent figures.

### 3.2. Plummer spheres with increasing inclination

The magnetic field patterns in Figure 1 are calculated assuming that the mean magnetic field direction lies in the plane of the sky. They show that the pinch of the pattern increases with increasing peak density, as expected if each field line is pulled inward as the density increases.

The shape of the pattern also depends on the inclination $\theta$ of the magnetic axis from the plane of the sky toward the line of sight, as seen in the simulations of K12 and also in Figure 2 of this paper. The plots in Figure 2 were calculated for a Plummer sphere with $\nu_0 = 300$, as in the center panel of Figure 1, using equations (21) - (24) and the same two azimuth planes $\phi = 0$ to $\pi$ and $\phi = \pi/2$ to $3\pi/2$. As $\theta$ increases, the projected height of a given point above its midplane decreases, while its projected cylindrical radius is unchanged. Thus for the range of $\theta$ where the projected flux tube shape resembles an hourglass, the pinch ratio $x_{max}/x_{min}$ does not change with increasing $\theta$.

However as $\theta$ increases, the hourglass shape becomes increasingly foreshortened, and field lines near the center become more radial and less vertical. As the magnetic axis direction approaches the line of sight, the hourglass is seen pole-on. Then its projected field lines approach a purely radial pattern extending from the minimum to the maximum flux tube radius. The radial lines do not pass through the origin, but have a gap of two scale lengths, corresponding to the waist of the smallest flux tube modelled.

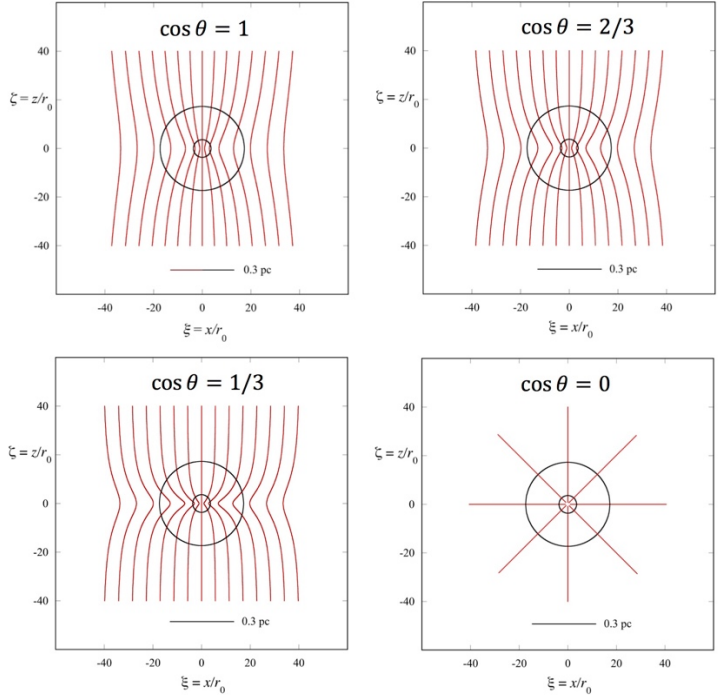

**Figure 2.** Projected flux tube lines for four inclinations $\theta$ of the magnetic axis from the plane of the sky toward the line of sight. The field lines thread a $p = 2$ Plummer sphere whose peak density exceeds the background density by a factor of 300. The 0.3 pc scale bar is based on background density $n_u = 300 \text{ cm}^{-3}$. The inner circle marks a density less than the peak value by a factor of 14.0, the same factor as in a critical Bonnor-Ebert sphere. The outer circle marks a density with twice the background value. For each of the four panels, the flux tube lines were calculated in the azimuthal planes $\phi = 0$ to $\pi$ (plane of sky) and in the perpendicular plane $\phi = \pi/2$ to $3\pi/2$ before inclination. For the fourth panel (pole-on view), additional flux tube lines were calculated in the azimuthal planes $\phi = -\pi/4$ to $3\pi/4$ and $\phi = \pi/4$ to $5\pi/4$ before inclination, to better illustrate the purely radial pattern of flux tube lines for this viewing orientation.

To test the choice of the planes used to produce the flux tube patterns in Figure 2, patterns were also calculated from equations (20) - (21) using a pair of azimuth planes rotated from the previous two planes by $-\pi/4$, i.e. planes with $\phi = -\pi/4$ to $3\pi/4$ and $\phi = \pi/4$ to $5\pi/4$. At $\theta = 0$ the resulting patterns (not shown) have the same hourglass shape as before, with horizontal extent reduced by the factor $\cos \pi/4$. At $\theta = \pi/2$ the pattern has four radial arms as in the previous case, but the phase of this radial pattern is shifted from its previous phase by $\pi/4$, as is shown in the fourth panel of Figure 2.

At intermediate inclinations near $\theta = \pi/4$, the analytic pattern arising from these rotated azimuth planes differs significantly from the pattern based on the original planes. Now the pattern is a superposition of two vertically displaced hourglass shapes arising from the front and rear portions of the flux tube. This front-back asymmetry is also seen in K12. In this K12 simulation, the difference in projected field direction between the front and back of each inclined flux tube decreases the polarized intensity in equatorial zones, as explained in K12 Figure 7. Nonetheless the predicted net polarization pattern in such regions of front-back asymmetry retains the hourglass shape, as seen in K12 Figure 6.

A similar effect occurs in the analytic model if the vertically displaced hourglass patterns are averaged together. Then the front and back deviations cancel and the net pattern matches the hourglass shape of the original planes. Thus, simple averaging of analytic profiles displaced by front-back asymmetry may provide a first approximation to the simulated field line directions arising from more detailed integration along the line of sight. On the other hand, accurate prediction of the polarized intensity, and its decrease due to front-back asymmetry, requires detailed density-weighted integration of the field components along the line of sight. This does not appear possible with SFF evaluation of field directions in individual planes.

### 3.3. Plummer prolate spheroids with increasing peak density

Filamentary structure is a basic property of star-forming clouds (Arzoumanian 2016). It is increasingly important to understand the processes that determine the direction of filament elongation relative to the polarization direction of the associated magnetic field (Myers & Goodman 1991, Goodman et al. 1992, Crutcher 2012, H.-B. Li et al. 2014, Auddy et al. 2016, Soler et al. 2016, Ade et al. 2018). Some studies have found that the distribution of such angles is not random, but rather that polarization directions tend to lie along the axes of low-density filaments and across the axes of high-density filaments (Soler et al. 2016). Here flux-freezing patterns are presented for prolate spheroids which are elongated perpendicular and parallel to the magnetic axis.

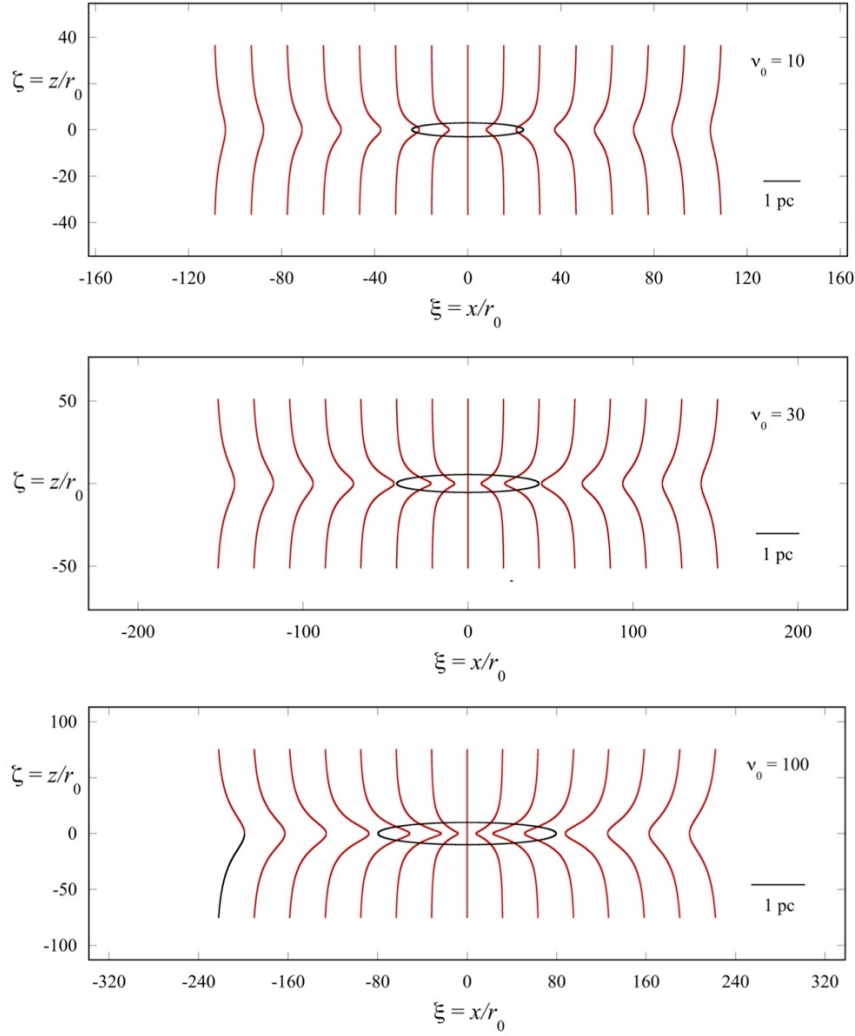

**Figure 3.** Flux tube lines for prolate Plummer spheroids having aspect ratio 8, ratio of peak density to background density 10, 30, and 100, and symmetry axis perpendicular to the magnetic axis, which lies in the plane of the sky. The ellipses mark a density twice the background density, as in Figures 1 and 2.

3.3.1. *Prolate spheroids perpendicular to the magnetic axis.* The case of prolate spheroids perpendicular to the magnetic axis has geometry similar to that of the magnetic ribbon model studied by Auddy et al. (2016); see also Tomisaka (2014) and Hanawa & Tomisaka (2015). Figure 3 shows flux tube patterns for perpendicular prolate spheroids having aspect ratio $A = 8$ and peak density ratio $\nu_0 = 10$, 30, and 100. The method of calculation and display is essentially the same as for the Plummer spheres in Figure 1. The peak density ratios are chosen to be less than in Figure 1, to more closely match observed patterns, and to reflect the property that filament gas is generally less dense than that of its embedded cores.

Figure 3 shows patterns which resemble those for spheres in Figure 1, which have been stretched in the $x$ - direction. They appear to have the same pinch ratio for the same density ratio and flux tube. For example, when $\nu_0 = 30$ the third flux tube has its waist at twice the background density, and its pinch ratio is $x_{max}/x_{min} = 1.5$ in both Figure 1 and Figure 3. However, for this flux tube the central field line directions differ: they are more radial for the prolate spheroid, and more vertical for the sphere. This shape difference resembles that between a sphere with high and low inclination angles between the plane of the sky and the observer, as shown in the third and first panels in Figure 2.

3.3.2. Prolate spheroids parallel to the magnetic axis. Figure 4 shows field line patterns associated with prolate spheroids which differ from those in Section 3.3.1 only in that their orientation is parallel to the magnetic axis, rather than perpendicular.

Figure 4 shows that the polarization signature for an aligned prolate spheroid is confined to a much more limited spatial zone than for a perpendicular prolate spheroid. The pinch of field lines is discernable only for a few scale lengths $r_0$ in the $x$ - direction, and for a few $Ar_0$ in the $z$ - direction, and only for $\nu_0 \gtrsim 30$. This property appears to arise because unlike the other cases considered, the elongation of this parallel prolate spheroid stretches the hourglass pinch along the hourglass axis. Detection of this signature would seem to require polarization observations which can sample the densest parts of the filament, rather than large-scale observations of the filament environment.

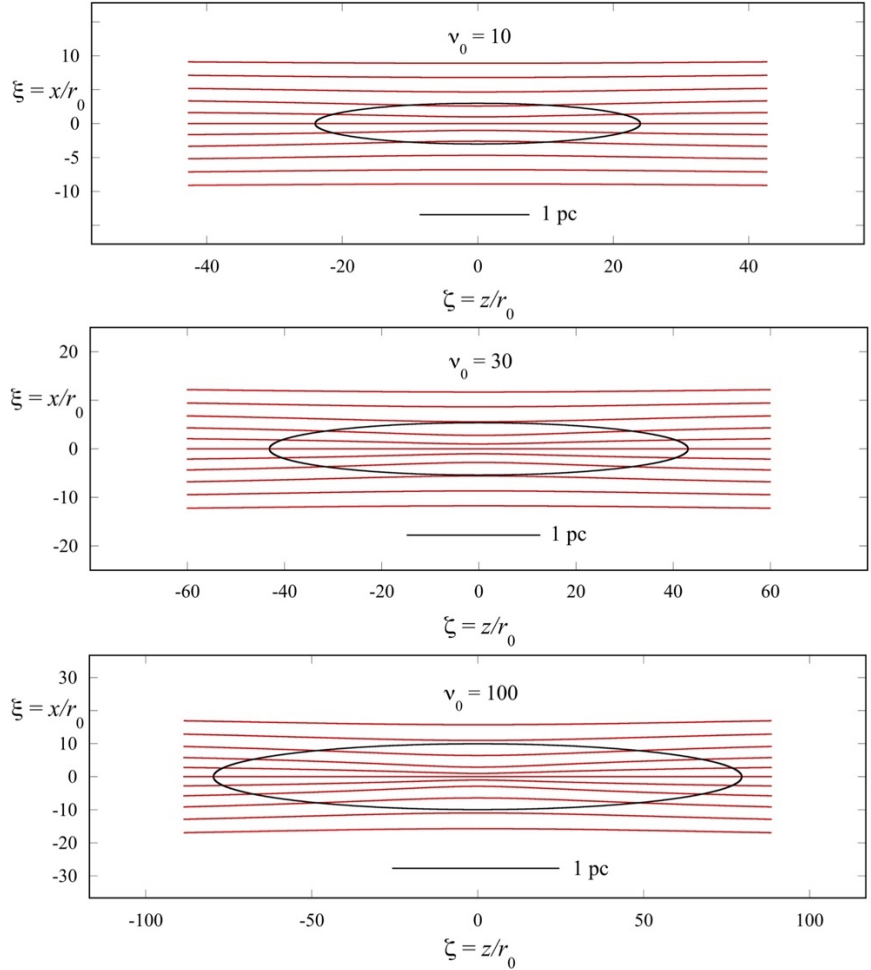

**Figure 4.** Flux tube lines for prolate Plummer spheroids having aspect ratio 8, ratio of peak density to background density 10, 30, and 100, and symmetry axis parallel to the magnetic axis, which lies in the plane of the sky. The ellipses mark a density twice the background density, as in Figures 1 - 3. These spheroids are the same as in Figure 3, except that they are rotated by 90 degrees, so that they are parallel rather than perpendicular to the direction of the mean magnetic field.

### 3.4. Oblate Plummer spheroids with increasing inclination

Flux freezing may apply to regions compressed by winds and supersonic flows associated with H II regions. Flux freezing may also apply to the outer zones of protostellar disks, where scattering does not dominate the polarization, and where non-ideal MHD processes do not dominate the magnetic field structure (Z.-Y. Li et al. 2014, Masson et al. 2016). Simulations comparing magnetic field structure in the $x$ - $z$ plane for ideal and nonideal MHD at the time of second core formation show similar hourglass shape for radii greater than $\sim 20$ au (Dapp et al. 2012, Vaytet et al. 2018).

In this section the magnetic field structure and a constant-density contour are presented for a flux-frozen oblate spheroid having three different inclinations. The variation with inclination of a field frozen into a flattened spheroid is instructive since the effect of the inclination is seen in both the field line pattern and in the projected shape of the spheroid.

Figure 5 shows the effect of increasing inclination on the field line pattern of an oblate Plummer spheroid with peak density ratio 300, aspect ratio 8, and symmetry axis parallel to the magnetic axis, as its symmetry and magnetic axes are inclined from their initial position in the plane of the sky, with inclination factor $\cos\theta$ varying from 1 to 2/3 to 1/3.

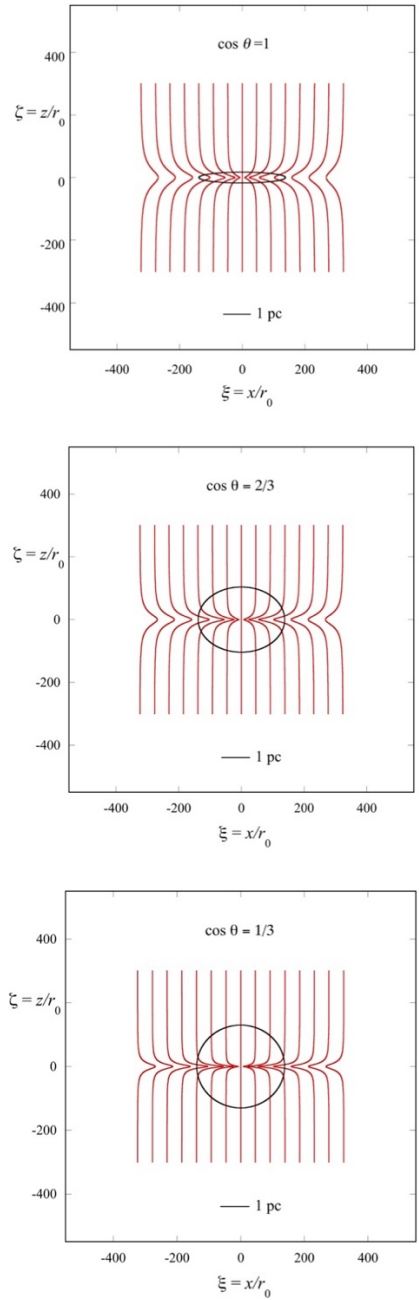

**Figure 5.** Projected flux tube lines for an oblate Plummer spheroid having aspect ratio 8, ratio of peak density to background density 300, and inclination factor $\cos\theta$ varying from 1 to 2/3 to 1/3. The ellipses mark a density twice the background density, as in Figures 1 - 4.

As with the inclined sphere, the maximum and minimum cylindrical radii of a flux tube do not change with inclination. As with the inclined sphere, the field lines near the center become more radial and less vertical with increasing inclination. However, in contrast to the inclined sphere, a surface of constant density in an oblate spheroid changes its projected shape with inclination, from a flattened ellipse having initial aspect ratio $A > 1$ at $\theta = 0$ to a circle having aspect ratio = 1 at $\theta = \pi/2$. The aspect ratio $A_{os}(\theta)$ of this ellipsoidal surface of a projected oblate spheroid can be written

$$A_{os}(\theta) = \{(1/2)[1 + A^{-2} - (1 - A^{-2})\cos 2\theta]\}^{-1/2} \qquad (27)$$

based on the maximum height of an inclined ellipse. If an observed condensation has an ellipsoidal projected shape, equation (27) can give the inclination angle needed to match the observed shape with an inclined oblate spheroid model. In turn this inclination angle can be used to model the shape of the associated inclined flux tube.

The inclined oblate spheroids in Figure 5 have field line patterns with significantly greater pinch than do the inclined spheres in Figure 2, having the same density contrast and the same inclination. This property may be understood if each oblate spheroid is considered to be a compressed version of the corresponding sphere, where the flux-frozen field lines have been dragged toward its equatorial plane during its compression.

### 3.5. Comparison with polarization simulation

This section quantifies the match between the field line and polarization patterns expected for a simple source with known properties of density, field strength, and mass-to-flux ratio. For this purpose the most relevant comparison is between the SFF model and the simulation study of K12.

K12 predict the expected polarization structure of a BE sphere threaded by a uniform magnetic field, as it collapses to form a protostar. The polarization is obtained from density-weighted integration along the line of sight of the horizontal and vertical magnetic field components, assuming optically thin emission by magnetically aligned dust grains (Tomisaka 2011). The initial state of the K12 model 1 is a critically stable BE sphere of mass 1.1 $M_\odot$, radius 6500 au, and temperature 10 K. It is threaded by a magnetic field having strength 23 $\mu$G and vertical direction in the plane of the sky, with a mass-to-flux ratio 6.6 times the critical value. It is thus a "weak-field" system where the sphere's initial magnetic energy is 6% of its initial gravitational potential energy. A polarization map of this model at the time 7 $10^4$ yr, or about two free-fall times since the start of collapse, just prior to formation of the protostar, is given in K12 Figure 5b. In each quadrant of the map, 64 polarization directions are plotted on a rectangular grid with spacing 400 au.

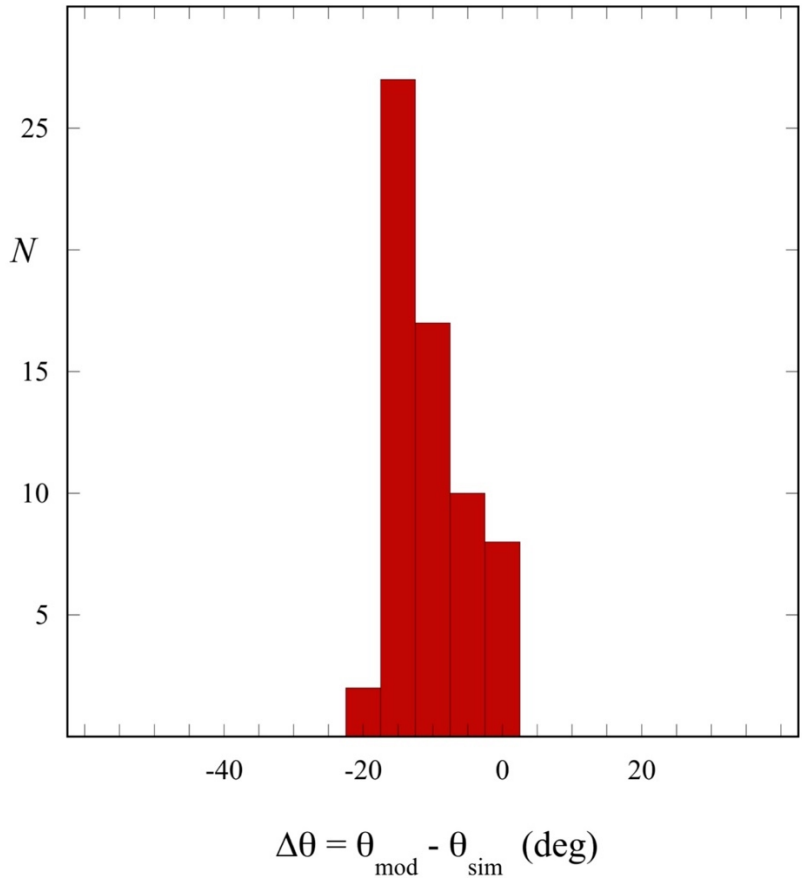

**Figure 6.** Histogram of relative orientations for SFF plane-of-sky field directions and K12 polarization directions for each of 64 map positions in the centrally condensed, magnetized sphere of K12 Figure 5b.

The K12 and SFF angle direction maps have similar hourglass shape, but the K12 polarization map has a weaker central pinch than the SFF field direction map. The relative orientations $\Delta\theta \equiv \theta_{mod} - \theta_{sim}$ have mean deviation $\overline{\Delta\theta}$ =-10 deg and standard deviation $\sigma_{\Delta\theta} =$ 6 deg, as shown in Figure 6. The maps differ (1) because the SFF model evaluates the field in the plane of the sky while the polarization simulation integrates the field along the line of sight, and (2) as the distance from the plane of the sky increases, the field direction approaches the background direction. Thus at each map position, the polarization direction slightly is more vertical than the plane-of-sky field direction.

The SFF model can approximate the line-of-sight integration of polarization observations by evaluating the field direction in a plane which is offset from the plane of the sky, and/or by evaluating the field direction in a plane at map positions $(\xi, \zeta)$ which have been "stretched" to positions $(\xi_s, \zeta_s)$, where $\xi_s > \xi$ and $\zeta_s > \zeta$. To quantify this approximation, the field direction was calculated from equations (5), (6), and (26) for positions with squared dimensionless spherical radius $\omega^2 = \eta_0^2 + \alpha^2(\xi^2 + \zeta^2)$. Here $\eta_0$ is a fixed offset of the evaluation plane from the plane of the sky, and $\alpha$ is a radial stretch factor within the evaluation plane.

The parameters $\eta_0$ and $\alpha$ were varied to minimize the mean and standard deviation of the 64 angle differences between the SFF field directions and the K12 polarization directions. It was found that $\eta_0 = 200$ and $\alpha = 2.34$ give $\overline{\Delta\theta} = 0.0$ deg and $\sigma_{\Delta\theta} = 1.3$ deg as shown in Figure 7. Here $\eta_0 = 200$ corresponds to moving the evaluation plane by a line-of-sight distance of 2900 au from the plane of the sky, or by 0.45 of the initial BE sphere radius. This simple modification of the SFF evaluation plane matches the polarization directions in K12 Figure 5b to within the estimated measurement uncertainty of ~1 deg.

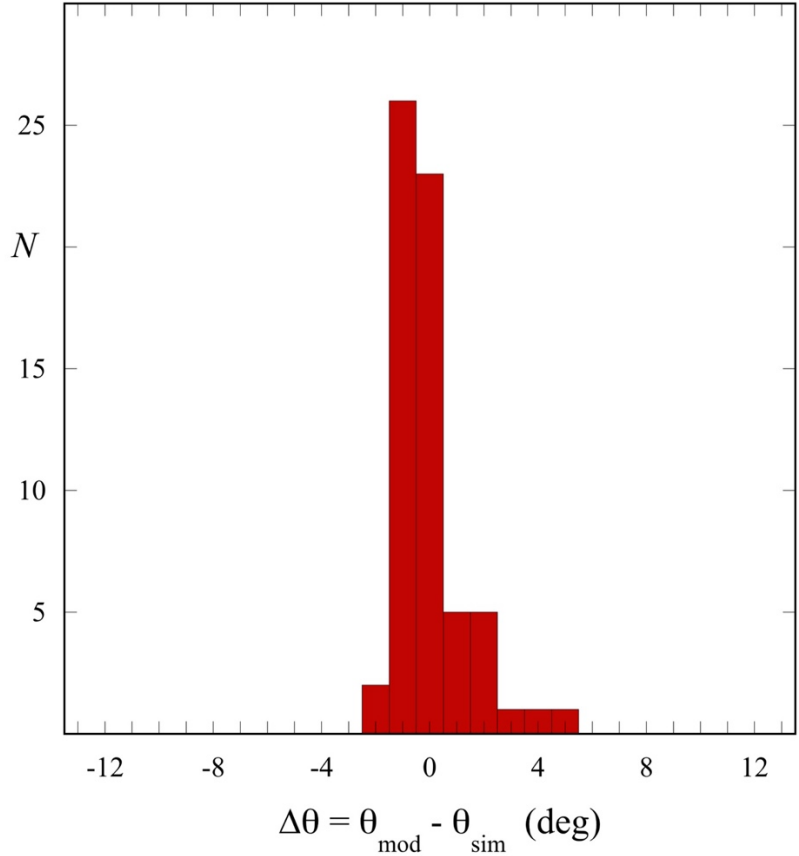

**Figure 7.** Histogram of relative orientations for SFF field directions in a modified evaluation plane and K12 polarization directions, for each of the 64 map positions as in Figure 6, in the centrally condensed, magnetized sphere of K12 Figure 5b. The SFF field directions in this plane match the K12 polarization directions more closely than do the plane-of-the-sky directions used in Figure 6.

These results indicate that the SFF model of magnetic field directions in a single plane can approximate an hourglass map of submillimeter polarization directions expected from a detailed numerical simulation, for a spheroidal condensation embedded in an initial background of uniform density and field strength. With no correction for line-of-sight integration, half the deviations are less than $\sim$ 10 deg, since the histogram of relative orientations has mean and standard deviation ( $\overline{\Delta\theta}$, $\sigma_{\Delta\theta}$ ) = (-10 deg, 5.6 deg) as shown in Figure 6. With correction for line-of-sight integration, half the deviations are less than $\sim 1$ deg, since $(\overline{\Delta\theta},\ \sigma_{\Delta\theta}) = (0.0\ \mathrm{deg},\ 1.3\ \mathrm{deg})$ as shown in Figure 7.

### 3.6. Comparison of SFF pattern with an observed polarization map

The fit of a SFF model to an observed polarization map is illustrated in Figure 8, based on ALMA observations of 1 mm polarization in the extended disk around the protostar VLA 1623A (Sadavoy et al. 2018, Figure 9). The SFF model was constructed by assuming that the magnetic field threads an oblate spheroidal disk. The disk symmetry axis is tilted toward the observer so that its projected shape matches the ellipsoidal shape of the extended emission. The tilted field pattern was computed as described in Sections 2.4 and 3.4. The SFF pattern was translated and rotated on the plane of the sky so that its symmetry axis passes near the center of the protostar emission, and so that the flux contours match the directions of the local polarization segments. The fit was optimized by eye. It is capable of further optimization, to better constrain parameter values and uncertainties.

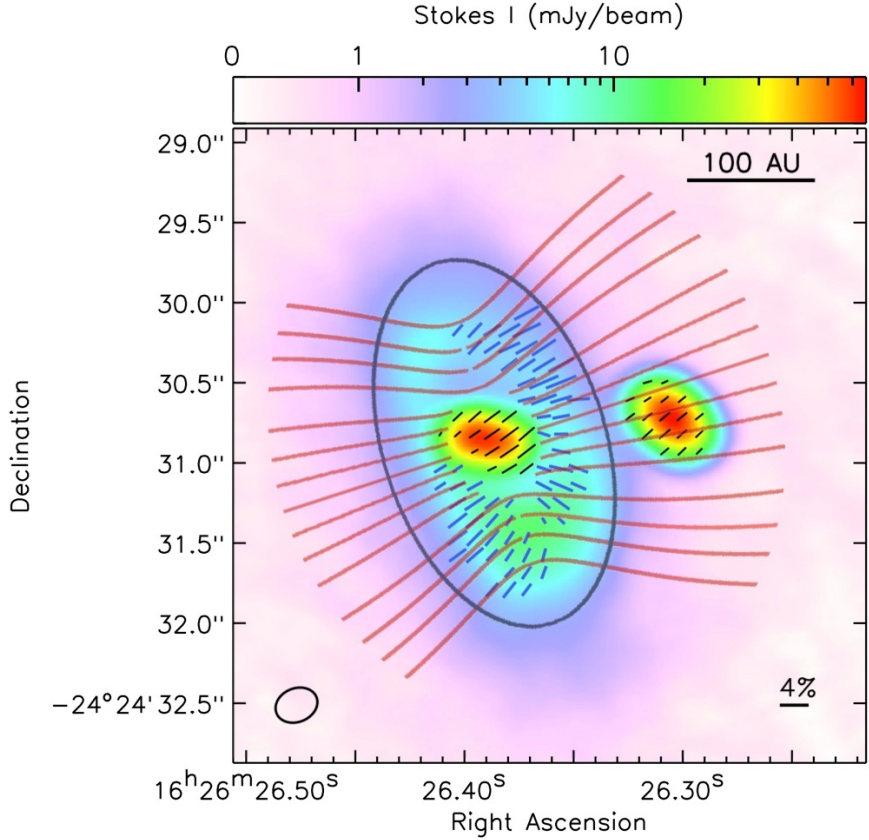

**Figure 8.** ALMA 1.3 mm flux and polarization image of VLA 1623 protostars A and B, the extended disk around VLA 1623A, and SFF model magnetic field lines threading a tilted model spheroid. Black segments are e-vectors, blue segments are b-vectors. The black ellipse shows the outline of the model spheroid. From Sadavoy et al (2018).

The observed polarization directions in Figure 8 can be classified by their locations in the map quadrants, which are here denoted as N (from NE to NW), E (from NW to SW), S (from SW to SE), and W (from SE to NE). The polarization directions agree well with the flux contours in the N and S portions of the map, where the polarization directions have approximate line symmetry about an axis through the protostar VLA 1623A, and less well in the W region, where the polarization directions are highly ordered, but depart from line symmetry. The W directions resemble those in an inclined rotating system after protostar formation, as shown in Figure 10b of K12. The SFF model field directions and observed polarization directions are compared in Figure 9, which shows the histogram of relative orientations $\Delta\theta \equiv \theta_{mod} - \theta_{obs}$ for all points with polarized intensity greater than 5-sigma, for the entire map and for the N-S region. For the entire map the mean and standard deviation are $(\overline{\Delta\theta}, \sigma_{\Delta\theta}) = (2 \text{ deg}, 19 \text{ deg})$; for the N-S region they are (0 deg, 9 deg).

If the deviation of 9 deg is due mainly to fitting errors, it is expected that the goodness of fit will improve when the SFF model is fit to the observed polarization map with statistically optimized parameters. If instead the deviation is due mainly to turbulent motions which drive Alfvenic fluctuations, the DCF method can be used to estimate the associated mean magnetic field strength. Assuming the temperature and mean density values adopted by Sadavoy et al. (2018), that the nonthermal velocity dispersion is equal to the thermal velocity dispersion, and that the DCF correction coefficient is equal to 1/2 (Ostriker et al. 2001), the mean field strength over the ~100 au extent of the region is ~ 0.06 G. This value lies within the range computed for a protostellar disk at the time of second core formation, in the region where ideal and non-ideal MHD field strengths overlap, according to the simulation of Vaytet et al. (2018).

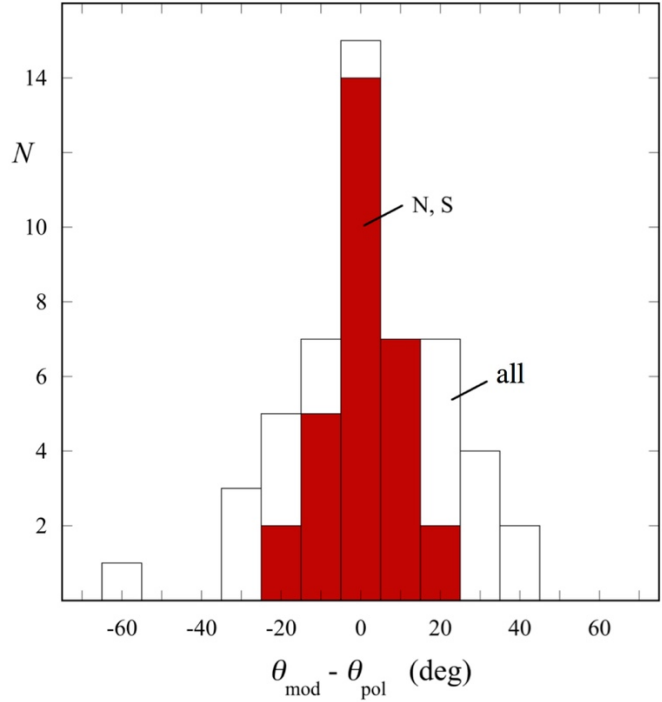

**Figure 9**. Histogram of relative orientations between SFF magnetic field directions projected on the plane of the sky and observed polarization directions in VLA 1623 as shown in Figure 8, for 51 points with polarized intensity greater than 5-sigma. For all 51 points (*white bars*) the distribution of relative orientations has mean and standard deviation 2 deg and 19 deg. For the 30 N-S points whose polarizations have approximate line symmetry (*red bars*) the mean and standard deviation are 0 deg and 9 deg.

These fit properties between the SFF model and observations in VLA 1623A are comparable to those for polarization models and observations in other nearby star-forming regions. In NGC1333 IRS 4A, models of magnetized accreting gas give $(\overline{\Delta\theta}, \ \sigma_{\Delta\theta}) = $ (-5 deg, 15 deg) for the pseudo-disk model of Galli & Shu (1993), and (-0.3 deg, 12 deg) for the Ohmic dissipation model of Shu et al. (2006) (Goncalves et al 2008). In BHB 07-11, the accreting and rotating disk model of Shu et al. (2007) fits best with a ratio of toroidal and poloidal field strengths = 1/3, giving $(\overline{\Delta\theta}, \ \sigma_{\Delta\theta}) = $ (23 deg, 7.5 deg) (Alves et al. 2018).

## 4. Inferring magnetic field strength from a polarization map

The flux freezing model described in Sections 2-3 predicts maps of the direction of the magnetic field associated with a variety of spheroidal condensations, for comparison with observed polarization maps. This section predicts complementary maps of the associated magnetic field strength for condensations whose background density and field strength can be estimated.

### 4.1. Relative field strength

If the column density map of a condensation can be described by a simple spheroidal density model, and if its polarization map matches the pattern expected from flux freezing in the same spheroid, the associated map of relative field strength follows directly from the model parameters.

The dimensionless field strength $\beta \equiv B/B_u$ at each point with cylindrical radius $\xi_c$ and spheroid radius $\omega$ is obtained from the quadrature sum of the radial and polar field components. These components are calculated from derivatives of the flux, e.g. in equations (11) and (12) of M66. Then the field strength at each position can be expressed in terms of the spheroid density $\nu$ in equation (5), the mean density $\bar{\nu}$ in equation (6), and the spheroid coordinates as

$$\beta = \bar{\nu}^{2/3} \left\{ 1 - \left(\frac{\xi_c}{\omega}\right)^2 \left[ 1 - \left(\frac{\nu}{\bar{\nu}}\right)^2 \right] \right\}^{1/2} . \qquad (28)$$

Equation (28) verifies that the field approaches the uniform background value at large radius, i.e. $\beta \to 1$ when $\nu \to 1$ and $\bar{\nu} \to 1$. Equation (28) also reproduces the peak value $\beta \to (1 + \nu_0)^{2/3}$ as $\omega \to 0$ (M66). Along the principal axes in the $x$-$z$ plane, the expression for the field strength simplifies from equation (28) to $\beta(x_c; z = 0) = \nu/\bar{\nu}^{1/3}$ and $\beta(z; x_c = 0) = \bar{\nu}^{2/3}$.

Contours for constant field strength follow from equation (28) by solving for the cylindrical radius,

$$\xi_c = \omega \left[ \frac{1 - (\beta/\bar{\nu}^{2/3})^2}{1 - (\nu/\bar{\nu})^2} \right]^{1/2} \qquad (29)$$

provided $0 \leq \beta \leq \bar{\nu}^{2/3}$. The height $\zeta(\omega, \xi_c)$ follows from equation (14) for the sphere, or from equations (15) - (19) for other spheroid shapes and orientations. As in the calculation of contours of constant flux in Section 2.3, $\omega$ serves as a dummy variable in obtaining each relation between $\zeta$ and $\xi_c$.

Contours of constant field strength in the $x$ - $z$ plane for a Plummer sphere are shown in Figure 10, based on equations (28) and (29). They are superposed on contours of constant flux for the peak density $\nu_0 = 300$ as in the center panel of Figure 1. Each field strength contour resembles a closed figure formed from two vertically displaced circles. This noncircular contour shape arises because the radial and polar components of the field strength have slightly different dependence on density (M66). If the field components had the same scaling with density, these contours would in general be concentric ellipses, and in Figure 6 they would be concentric circles. The shape can be understood qualitatively as the superposition of a large-scale vertical field and a localized radial field which has been pulled inward by flux freezing during the condensation process.

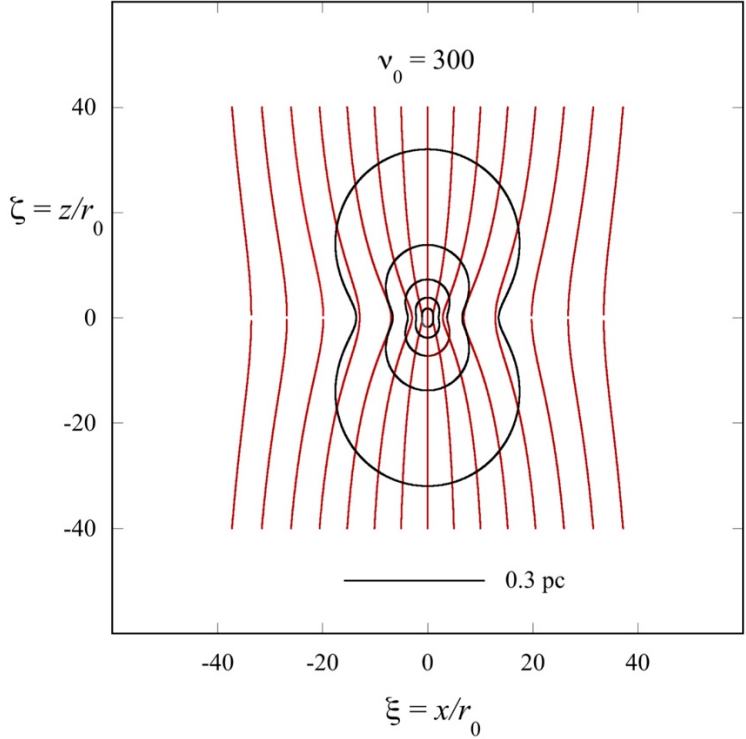

**Figure 10.** *Closed curves:* magnetic field strength which exceeds the background field strength by factors 1.5, 3, 6, 12, and 24. *Open curves:* flux tube lines as in Figure 1. The peak density exceeds the background density $n_0$ by a factor of 300. The scale bar is based on temperature 10 K and background density $n_u = 300 \, \text{cm}^{-3}$. With these parameters the scale length is $r_0 = 0.011$ pc.

In Figure 10, the equatorial pinch in the field strength reflects magnetic forces toward the equator which are not balanced by gravity or by a thermal pressure gradient. As the condensation contracts, these forces channel gas toward the equator, forming a non-equilibrium equatorial zone of enhanced density (M66). On protostellar scales of $\sim 100$ au, this zone has been called a "pseudo-disk" since it differs from a rotationally supported Keplerian disk (Galli & Shu 1993). Some simulations show rotational properties of Keplerian disks on smaller scales and magnetic properties of pseudo-disks on larger scales (Masson et al. 2016, Vaytet et al. 2018).

## 4.2. Background field strength and density

Estimation of the field strength from the above model requires knowledge of the peak density ratio $\nu_0$. a background field strength $B_u$, and a background density $n_u$. These background quantities can be considered large-scale-average properties of the gas from which the condensation formed. Zeeman observations indicate that the field strength scales with density as $B \propto n^\kappa$, $1/2 \lesssim \kappa \lesssim 2/3$, for $B \gtrsim 10$ $\mu$G and for $n \gtrsim 300$ cm$^{-3}$ (Crutcher 2012; Tritsis et al 2015). These values of $B_u$ and $n_u$ may be considered the lowest useful background values for flux-freezing models in nearby molecular clouds.

Estimates of background density for a particular cloud may be made by dividing the typical column density at the map boundary by its spatial extent. For example filament #6 in IC5146 has a background column density $\sim 1 \ 10^{21}$ cm$^{-2}$ extending for $\sim 1$ pc according to *Herschel* observations (Arzoumanian et al. 2011), indicating a mean background density $\sim 330$ cm$^{-3}$. For a large molecular cloud, where most of the mass is at the periphery, an upper limit on the mean background density is the mean density over the cloud itself. In five molecular clouds within 260 pc, defined by visual extinction greater than 2 mag, the median of their mean densities derived from star count extinction is 345 cm$^{-3}$ (Evans et al. 2009). Thus, a mean background density $\sim 300$ cm$^{-3}$ is plausible for these parsec-scale clouds.

## 4.3. Field strength estimate

Applying the foregoing properties to the contours in Section 4.1, if the peak density is $n_0 = 10^5$ cm$^{-3}$ in a region of background density $n_u = 330$ cm$^{-3}$ and background field strength $B_u = 10$ $\mu$G, the field strength contours in Figure 6 correspond to 15, 30, 60, 120, and 240 $\mu$G. Since $\nu_0 = n_0/n_u = 300$, the maximum field strength is $B_{\max} = B_u(1 + \nu_0)^{2/3} = 450$ $\mu$G.

A further constraint on the background density $n_u$ and field strength $B_u$ comes from the requirement that the mass $M$ within a given density contour exceed the magnetic critical mass, if magnetic forces are to allow gravitational collapse (MO07). The critical mass, defined as $M_c = \Phi/(2\pi G^{1/2})$, can be written in terms of $n_u$ and $B_u$ using equation (3) as

$$M_c = \left(\frac{3}{8\pi m n_u}\right)^2 \frac{B_u^3}{2G^{3/2}} \ . \qquad (30)$$

Thus if a truncated Plummer sphere of peak density $n_0 = 10^5$ cm$^{-3}$ condensed from a uniform medium with initial field strength $B_u = 10$ $\mu$G and initial density $n_u = 300$ cm$^{-3}$, the minimum mass it must have to allow collapse by self-gravity against its magnetic forces is 160 $M_\odot$ according to equation (30), a typical mass for a small embedded cluster. In contrast, if a truncated Plummer sphere of the same peak density condensed from the same initial field strength but from a greater initial density $n_u = 3700$ cm$^{-3}$, its minimum mass would be 1 $M_\odot$, typical for a dense core forming a low-mass star or a small multiple system. Magnetic critical masses for spheroidal geometry are given by Bertoldi & McKee (1992).

The magnetic critical mass in equation (30) can be considered analogous to the BE critical mass, since the two critical mass equations have exactly the same form, $M_c = cs^3 G^{-3/2} \rho^{-1/2}$. For the BE sphere, $s$ is the isothermal sound speed, $\rho$ is the boundary density and $c = c_{BE} = 1.182$ (MO07). For the magnetized sphere, $s$ is the background Alfven speed $B_u/\sqrt{4\pi\rho_u}$, $\rho$ is the background density $\rho_u$, and $c = c_{mag} = 9/(16\sqrt{\pi}) = 0.317$. Thus if a BE sphere and a magnetized sphere have the same critical mass and if the background Alfven speed is equal to the sound speed, the magnetic background density is less than the BE boundary density by a factor $\left(c_{BE}/c_{mag}\right)^2 = 13.9$.

## 5. Discussion

### 5.1. Summary

The model described in Sections 2-4 extends the analysis of M66, which was limited to spheres having a Gaussian density distribution. It assumes that a centrally concentrated spheroid condensed from an initial medium of uniform density and magnetic field, conserving mass, flux, and shape. It gives a scalable analytic way to predict the structure and strength of the magnetic field associated with a Plummer sphere, prolate spheroid, or oblate spheroid. The magnetic field structure is given as a function of the spheroid density contrast, aspect ratio, and orientation of the magnetic axis with respect to the line of sight. It includes cases where the spheroid symmetry axis is either parallel or perpendicular to the magnetic axis. The model can apply to observations of dense cores, filaments, and disks, whose ordered polarization is due to magnetically aligned grains, and whose structure is approximately axisymmetric.

The spheroid model parameters needed from observations are the peak density $n_0$ and thermal velocity dispersion $\sigma$, the background density $n_u$, and the aspect ratio $A$ for oblate or prolate spheroids. The background field strength $B_u$ is obtained from Zeeman observations if available, or from $B_u = 10 \ \mu$G, the minimum field strength which follows the scaling law expected for flux freezing (Crutcher 2012). The inclination $\theta$ of the magnetic axis from the plane of the sky toward the line of sight can be adjusted to improve the match to the observed column density map and hourglass shape.

The model can predict the polarization at every observed map point, to optimize the model parameters for best fit. The optimized parameters can then be used to compute a global field pattern at regularly spaced intervals on the plane of the sky.

The model can test the applicability of flux freezing compared to other field-gas interactions, including ambipolar diffusion and MHD turbulent motions. For example, cores formed from strongly supercritical conditions should have field lines with greater curvature, due to the dominance of field line dragging, while cores formed from marginally critical conditions should have less curvature, due to the greater role of ambipolar diffusion (Basu et al. 2009). It can estimate field strength for comparison to estimates from observations of the Zeeman effect, or from estimates based on fluctuations in polarization direction (Chandrasekhar & Fermi 1953), or from other methods described in Crutcher (2012).

## 5.2. Limitations

The SFF model in Sections 2-3 is highly idealized since its flux-frozen structures arise from a hypothetical initial medium which is uniform in density, field strength, and field direction. Although regions of nearly uniform polarization are seen around some molecular clouds, such simple cloud environments are rarely observed in column density maps. It may be sufficient to estimate an appropriate background density from an average over the column density map surrounding the region of interest, as discussed in Section 4.2.

The SFF model is also limited because its spheroid geometry restricts its application to condensations having simple axisymmetric shapes, without outflows, without multiple sources, without significant rotation, and without twisted fields, whose polarization patterns can be difficult to distinguish from those due to otherwise similar fields which are not twisted (Franzmann & Fiege 2017). Furthermore, the model applies only to ordered polarization patterns where the influence of turbulent field fluctuations can be neglected.

## 5.3. Uncertainties

An important feature of this model is the dependence of mean field strength $B$ on mean density $n$ as $B \propto n^{\kappa}$, $\kappa = 2/3$, as in M66. This exponent is similar to the best-fit value 0.65 for $n \geq 300$ cm$^{-3}$ found from fitting a composite function to Zeeman observations at all available densities (Crutcher 2012). However, an analysis of only the high-density Zeeman observation values finds an exponent closer to 0.5, as expected in moderate to strong field models (Tritsis et al. 2015). This range of exponents is also seen in a turbulent ideal MHD simulation, where strong fields yield $\kappa \approx 1/2$ and weak fields yield $\kappa \approx 2/3$ (Mocz et al. 2017).

Nonetheless the field strengths obtained from the present model do not change greatly between $\kappa = 1/2$ and $\kappa = 2/3$. For fixed background field strength, the peak field strength $B_0$ for $\kappa = 2/3$ exceeds that for $\kappa = 1/2$ by the factor $\nu_0^{1/6}$ where $\nu_0$ is the ratio of peak to background density. Thus, over the density range considered in this paper, increasing $\kappa$ from 1/2 to 2/3 means increasing the peak field strength by a factor 1.5 when $\nu_0 = 10$, up to a factor 2.6 when $\nu_0 = 300$.

The simple power-law relation between field strength and density due to flux freezing becomes more complex when non-ideal processes are considered. These include effects associated with higher densities, such as ambipolar diffusion (Ciolek & Mouschovias 1994), ohmic dissipation (Machida et al. 2006), and effects associated with greater turbulence, such as reconnection diffusion (Lazarian et al. 2014).

The model assumes shape-preserving contraction from a uniform initial spheroid. This idealization remains to be tested in detailed simulation studies. The spherical case is the least objectionable, since collapse from a spherical initial state at rest retains approximate spherical symmetry if the initial field is sufficiently weak. For nonspherical initial spheroids, it is assumed that magnetized, turbulent converging flows bring gas into filamentary or flattened structures, as is commonly seen in MHD simulations (e.g. Li et al. 2004, Lee & Hennebelle 2016). These initial configurations are assumed to contract further under self-gravity as their turbulence dissipates.

It seems unlikely that strict shape conservation can occur during evolution of a condensed structure, especially if turbulent motions are important in the evolution. However, the final properties of a condensation may not depend strongly on all of its initial properties. A detailed MHD study of simulated cores indicates similar final density profiles and mass-to-flux ratios, despite substantial differences in initial density, field strength, and ratio of turbulent to magnetic energy (Mocz et al. 2017).

## 5.4. Ways to infer magnetic field from polarization

Here the present SFF method of field estimation is compared to that of flux conservation in an hourglass flux tube (Schleuning 1998) and to the statistical inference of the mean field strength from the disordered component of the polarization pattern (DCF).

Hourglass polarization patterns have been used to estimate relative field strengths based solely on flux conservation within a flux tube delineated by an hourglass polarization pattern. When the polarization pattern approximates an hourglass shape, as in submillimeter observations of Orion A, the ratio of hourglass maximum and minimum widths $\Delta x_{max}$ and $\Delta x_{min}$ at the broadest and narrowest widths of a flux tube is simply related to the mean field strengths within $\Delta x_{min}$ and $\Delta x_{max}$ by $\bar{B}(\Delta x_{min})/\bar{B}(\Delta x_{max}) = (\Delta x_{max}/\Delta x_{min})^2$. This estimate leads to a mean field ratio $\approx 4$ for the Orion A ridge, neglecting projection and optical depth effects (Schleuning 1998). In contrast, SFF applies flux freezing to the density structure of the entire spheroid model. As a result, SFF gives a 2D map of field strength and direction, while the hourglass width comparison gives the relative mean field strength at two positions in the map. On the other hand, SFF requires more parameters to specify the density model of the spheroid.

The DCF method applies to the fluctuating part of a polarization map due to turbulent excitation of Alfven waves, unlike the SFF method, which applies to the ordered part of a polarization map, arising from flux freezing as the spheroid contracts. The DCF method requires knowledge of the dispersion in polarization directions, of the mean gas density in the region, and knowledge of the Alfven speed from the nonthermal component of an appropriate spectral line width. When the observed polarization has both ordered and fluctuating components it is necessary to separate them, to obtain the dispersion of the fluctuating component (Hildebrand et al. 2009, Kandori et al. 2017). The result of the DCF analysis is the mean field strength over a map region large enough to have a significant number of independent polarization measurements.

The SFF estimate of magnetic field strength has estimated systematic uncertainty of a factor of a few, due to uncertainty in background field strength and in the degree of magnetic criticality of the model spheroid. Its random uncertainty is limited by the quality of the model fit to the observed polarization pattern. Its spatial resolution depends on the resolution of the spheroid density model, which in turn is limited by the resolution of the column density observations. If the spheroid model fits the observations well, the effective resolution of the SFF estimate may be significantly finer than that of the corresponding DCF estimate.

## 5.5. SFF field structure with turbulent fluctuations

This section discusses possible consistency between SFF and DCF analysis. Since the SFF model requires a well-determined ordered component of polarization, the turbulent energy must be significantly less than the energy in the mean field. Nonetheless, it is useful to test whether SFF and DCF models are consistent when the polarization pattern of a region has evidence of an ordered and disordered component, and when the line width of a dense gas tracer has a nonthermal component attributable to Alfvénic motions.

Flux freezing in a spherical core relates the mean field strength $\bar{B}$ to the mean density $\bar{\rho}$, the mass $M$, and magnetic critical mass $M_c$, by

$$\bar{B}_{SFF} = \frac{8\pi G^{1/2} M_c R \bar{\rho}}{3M} \qquad (31)$$

since $M = 4\pi R^3 \bar{\rho}/3$, $M_c = \Phi/\left(2\pi G^{1/2}\right)$, and $\Phi = \pi R^2 \bar{B}_{FF}$. The DCF model relates $\bar{B}$ to $\bar{\rho}$ and to the ratio of the dispersion of nonthermal motions to the dispersion of polarization angles, $\sigma_{NT}/\sigma_\theta$, by

$$\bar{B}_{DCF} = \frac{1}{2}(4\pi\bar{\rho})^{1/2} \frac{\sigma_{NT}}{\sigma_\theta} \qquad (32)$$

where the DCF correction coefficient is assumed to be 1/2 (Ostriker et al. 2001). Equating these mean field strengths gives a simple expression for the expected polarization angle dispersion if the SFF and DCF field strengths are consistent,

$$\sigma_\theta = \frac{3}{8\pi} \frac{M}{M_c} \frac{\sigma_{NT}}{\sigma_T} \frac{\lambda_J}{R} \qquad , \qquad (33)$$

where $\lambda_J \equiv \sigma_T[\pi/(G\bar{\rho})]^{1/2}$ is the core Jeans length and $\sigma_T$ is the core thermal velocity dispersion.

To test consistency, equation (33) is applied to the well-studied starless core FeSt 1-457, for which the relevant properties are $M/M_c = 2.0$, $\sigma_{NT}/\sigma_T = 0.31$, and $\lambda_J/R = 1.6$ (Kandori et al. 2017; K17). The predicted polarization angle dispersion is then $\sigma_\theta = 6.8$ deg. The observed near-infrared polarization angle dispersion has an estimated upper limit $(\sigma_\theta)_{max} = 6.9 \pm 2.7$ deg (K17). This value is considered an upper limit because it is obtained by subtraction of a function which represents the ordered component of the pattern. For this function K17 used a parabola, rather than an hourglass description expected for flux freezing. If the observed dispersion is set equal to its upper limit, the observed and predicted dispersions agree within one-sigma uncertainty.

The agreement of $\sigma_\theta$ and $(\sigma_\theta)_{max}$ indicates that flux freezing and modest turbulent motions are consistent with the same mean field in FeSt 1-457, and with the observed degree of polarization fluctuations. On the other hand, greater levels of turbulent motion can be expected to distort the ordered component of the magnetic field, so that no simple pattern can be discerned, as in Ser-emb-8 (Hull et al. 2017). It remains to be studied how the SFF and DCF models can best be used when the observed polarization pattern has comparable contributions from its ordered and disordered components.

## 5.6. Future applications

The application of the SFF model to observations can be improved in concert with a suite of polarization simulations, where a family of source models of varying density, field strength, and orientation can be considered in greater detail.

In a region where the magnetic field has significant large-scale flux-freezing distortion and also significant small-scale fluctuations, each of the SFF and DCF methods is incomplete, because each uses one part of the available information and neglects the other part. It would be useful to compare results of the SFF and DCF methods on a sample of well-studied sources, as was done in Section 5.5 for FeSt 1-457.

It may also be useful include both the ordered and fluctuating component in the same model, to obtain more accurate estimates of field strength. An early effort in this direction was applied to distributions of optical polarization (Myers & Goodman 1991). Much more useful information on both polarization structure and density structure is now available, including well-resolved maps of column density from dust extinction (Lombardi et al. 2006) and emission (Andre et al. 2010).

Similarly, high-resolution observations of $NH_3$ line emission enable detailed maps of the nonthermal line width (Friesen et al. 2018), which may have a resolution advantage over polarization maps. They may therefore allow another estimate of field strength which is complementary to the SFF and DCF methods discussed here (e.g. Auddy et al. 2018).

It may be useful to extend the present SFF analysis of poloidal field structure to systems having both poloidal and toroidal field components, as is suggested by some ALMA observations of protostellar disks (e.g. Alves et al. 2018). It may be possible to approximate such a system as the projection of a poloidal field which has undergone a simple twist about its axis.

Large-scale cloud surveys such as BISTRO (Ward-Thompson et al. 2017) of cloud complexes which include filaments, cores, and hubs may provide useful application of the SFF model, to study how polarization structure varies with the geometry of the associated dense gas.

## 6. Conclusion

The main points of this paper are:

1. An analytic, scalable model of magnetic field direction and strength is based on flux freezing in condensations of spheroidal shape, including prolate and oblate spheroids of varying orientation.

2. The model extends the treatment of Mestel (1966) from spheres to spheroids. Each spheroid has a Plummer density profile of index $p = 2$. The model parameters are the background density and field strength, the peak density, the spheroid aspect ratio, and the inclination of the magnetic axis from the plane of the sky.

3.    The field energy is weaker than the gravitational energy, enabling the spheroid to collapse and form stars.  The field ordered component is stronger than its turbulent component, so that field lines can have large-scale hourglass shape with no fluctuations.

4.    The model can test the prevalence of flux freezing in centrally condensed clouds whose polarization maps approximate an hourglass shape.  It can apply to polarization maps of dense cores, filamentary clouds, and circumstellar disks, where the field structure is dominated by flux freezing and where polarization due to magnetically aligned grains is not confused by other mechanisms of polarization.  For a given spheroid, the model provides global patterns of polarization in the plane of the sky as contours of constant flux, and local values of polarization angle at any map point.

5.    The hourglass shape of a flux tube depends on its flux, on the ratio of peak to background density, on the shape of the spheroid, and on its inclination.   At large distance from the center, the field lines approach the parallel and uniformly spaced vertical configuration of the original medium.  At distances within a few spheroid scale lengths, the field lines converge in a "pinch" with a nearly radial distribution.

6.  The pinch becomes more pronounced with increasing density ratio and with increasing concentration of dense gas in the equatorial plane.  For all spheroid shapes, the pinch increases with increasing density ratio as in Figure 1.  For a given density ratio, the pinch of the spheroid field pattern decreases if the spheroid shape is "stretched" in the direction of the magnetic axis, as in the prolate spheroid in Figure 3.  Similarly, the pinch increases if the spheroid shape is "squeezed" along the magnetic axis, as in the oblate spheroid in Figure 5.

7.  The field line directions in the pinch region become more radial and less vertical with increasing inclination of the magnetic axis from the plane of the sky.   In the limit of maximum inclination, where the magnetic axis lies along the line of sight, the field line directions become perfectly radial, as in Figure 2.

8.  For polarization maps well-fit by the model, the derived map of field strength can be compared with field strength estimates based on the fluctuating field component as in the DCF method.  The model outputs can also serve as inputs to a polarization radiative transfer code for more accurate predictions of polarization and field structure.

9.  SFF magnetic field directions typically agree within ~ 10 deg of the corresponding polarization directions in a simulation of a collapsing BE sphere, over 64 map points.  The deviation is due mainly to SFF evaluation in the plane of the sky and to integration along the line of sight to compute polarization.  When the position and scale of the SFF evaluation plane are increased to minimize deviations, the typical deviation is reduced to ~ 1 deg.

10.  A SFF model fits ALMA polarization directions in VLA 1623A with 1-sigma deviation of 9 deg for 30 map points having approximate line symmetry and 19 deg for all 51 map points having polarization signal-to-noise ratio > 5 (Sadavoy et al. 2018).  These deviations are comparable to those of polarization model fits in NGC 1333 IRS4 (Gonsalves et al. 2008) and in BHB 07-11 (Alves et al. 2018).

11.  The mean magnetic field strength $\bar{B}_{SFF}$ based on SFF is consistent with $\bar{B}_{DCF}$ based on Davis-Chandrasekhar-Fermi analysis of polarization dispersion, density, and turbulent motions in the starless core FeSt 1-457 (Kandori et al. 2017; K17).  Assuming $\bar{B}_{SFF} = \bar{B}_{DCF}$ yields a predicted angle dispersion 6.8 deg.  This dispersion agrees within uncertainty with the observed dispersion, which has an upper limit of $6.9 \pm 2.7$ deg (K17).

**Appendix**

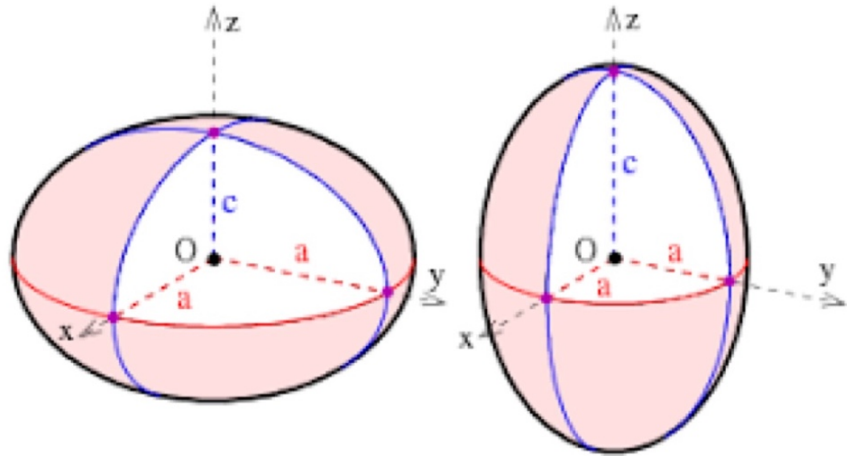

**Figure A1.** Oblate spheroid (*left*) and prolate spheroid (*right*), showing that in each case a planar cut perpendicular to the symmetry axis (*z*-axis) is a circle *(red line)*. In each case a planar cut parallel to the symmetry axis is an ellipse *(blue line)*. The orientation of each ellipse matches that of its spheroid. For the oblate spheroid, the ellipse semi-minor axis is parallel to the short axis of the spheroid. For the prolate spheroid, the ellipse semi-major axis is parallel to the long axis of the spheroid. Figure reference: https://en.wikipedia.org/wiki/Spheroid

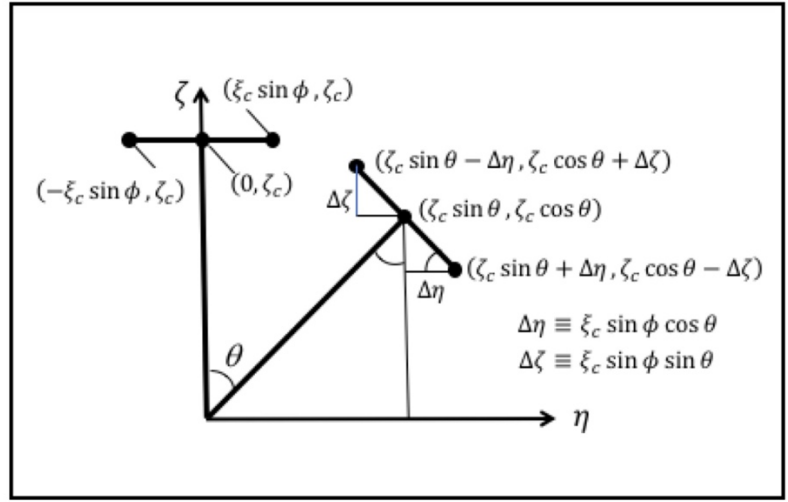

**Figure A2.** Coordinates of points on a tilted flux tube, useful in deriving equations (20) and (21). The figure shows the projection onto the $\eta - \zeta$ plane of a line along the diameter of a circular flux tube, at three times. (1) The initial line extends from $(\xi, \eta, \zeta) = (-\xi_c, 0, \zeta_c)$ to $(\xi_c, 0, \zeta_c)$, and it appears in projection at the position $(\eta, \zeta) = (0, \zeta_c)$. (2) After rotation through azimuth angle $\phi$ about the $\zeta -$ axis, the line extends from $(\eta, \zeta) = (-\xi_c \sin\phi, \zeta_c)$ to $(\xi_c \sin\phi, \zeta_c)$. In the $\xi -$direction this line extends to $\xi_c \cos\phi$, equal to $\xi_{\theta\phi}$ in equation (20). (3) After inclination through polar angle $\theta$ about the $\xi -$ axis, the line extends from $(\eta, \zeta) = (\zeta_c \sin\theta - \Delta\eta, \zeta_c \cos\theta + \Delta\zeta)$ to $(\zeta_c \sin\theta + \Delta\eta, \zeta_c \cos\theta - \Delta\zeta)$, where $\Delta\eta$ and $\Delta\zeta$ are defined in the figure. These coordinates may be verified using the property that each angle indicated by an arc has the same value $\theta$. The height $\zeta_c \cos\theta - \Delta\zeta$ of the lower right point is equal to $\zeta_{\theta\phi}$ in equation (21).

**Acknowledgements**

Discussions with Joao Alves, Tyler Bourke, Blakesley Burkhart, Mike Dunham, Alyssa Goodman, Alex Lazarian, Thushara Pillai, Riwaj Pokhrel, Sarah Sadavoy, Ian Stephens, Enrique Vazquez-Semadeni, and Qizhou Zhang are gratefully acknowledged. The comments and suggestions of the referee, Robi Banerjee, and of an anonymous referee, led to significant improvements in the paper.